\documentclass[aps,prd,amsmath,amssymb,
twocolumn,
floatfix
]{revtex4-1}
\usepackage{graphicx}
\usepackage{times}
\usepackage{multirow}
\usepackage{verbatim}
\usepackage{color,soul}
\usepackage{xcolor}
\usepackage{cancel}
\usepackage{hyperref}
\usepackage[bottom]{footmisc}
\usepackage[normalem]{ulem}
\usepackage{fontawesome5}

\graphicspath{{figures/}}

\usepackage[draft,inline,nomargin]{fixme}
\fxsetup{theme=color, mode=multiuser}
\FXRegisterAuthor{pts}{1}{\color{red}PTS}
\FXRegisterAuthor{add}{2}{\color{blue}Addition}

\newcommand{\nuebar}{\ensuremath{\overline{\nu}_{e}} }
\newcommand{\uFive}{$^{235}$U}
\newcommand{\uEight}{$^{238}$U}
\newcommand{\pNine}{$^{239}$Pu}
\newcommand{\pOne}{$^{241}$Pu}
\newcommand{\pZero}{$^{240}$Pu}
\newcommand{\xSec}{\ensuremath{\times10^{-43}\,\text{cm}^2 / \text{fission} }}

\makeatletter
\newcommand{\github}[1]{%
   \href{#1}{\faGithubSquare}%
}
\makeatother

\begin{document}

\title{Exploring Current Constraints on Antineutrino Production by \pOne~and Paths Towards the Precision Reactor Flux Era}

\author{Yoshi Fujikake}%
\author{Bryce Littlejohn}
\email{blittlej@iit.edu}
\author{Ohana B. Rodrigues}%
\email{obenevidesrodrigues@iit.edu}

\affiliation{Physics Department, Illinois Institute of Technology, Chicago, IL 60616, USA}%

\author{Pranava Teja Surukuchi}%
\email{pranavateja.surukuchi@yale.edu}

\affiliation{Wright Laboratory, Yale University, New Haven, CT 06520, USA}%

\begin{abstract}
By performing global fits to inverse beta decay (IBD) yield measurements from existing neutrino experiments based at highly $^{235}$U enriched reactor cores and conventional low-enriched cores, we explore current direct bounds on neutrino production by the sub-dominant fission isotope $^{241}$Pu.  For this nuclide, we determine an IBD yield of $\sigma_{241}$ = 8.16 $\pm$ 3.47 cm$^2$/fission, a value (135 $\pm$ 58)\% that of current beta conversion models.  This constraint is shown to derive from the non-linear relationship between burn-in of $^{241}$Pu and $^{239}$Pu in conventional reactor fuel.  By considering new hypothetical neutrino measurements at high-enriched, low-enriched, mixed-oxide, and fast reactor facilities, we investigate the feasible limits of future knowledge of IBD yields for \uFive, \uEight, \pNine, \pOne, and \pZero.  We find that first direct measurement of the \pZero~IBD yield can be performed at plutonium-burning fast reactors, while a suite of correlated measurements at multiple reactor types can achieve a precision in direct \uEight, \pNine, and \pOne~yield knowledge that meets or exceeds that of current theoretical predictions. 
\end{abstract}

\maketitle

\section{Introduction}\label{sec:introduction}

A nuclear reactor primarily generates thermal energy as product nuclei inherit (as kinetic energy) and deposit (through repeated elastic collisions) excess rest mass energy from the fission of heavy nuclides in the reactor's fuel, such as \uFive, \uEight, \pNine, \pOne, and more.   
Successive decays of these neutron-rich product nuclei release additional energy in the form of beta particles, gamma-rays, and antineutrinos.  
While the two former product types are additional, sub-dominant contributors to heat generation in a reactor, the antineutrinos (\nuebar) and their associated kinetic energy entirely escape the reactor core, offering an attractive avenue for studying the properties of neutrinos~\cite{reactor_review,qian_review,CONNIE:2022hna}, interrogating state-of-the-art nuclear data~\cite{bib:WoNDRAM}, and non-intrusively monitoring nuclear reactor cores~\cite{Bernstein:2019hix}.  
Reactor-based \nuebar detectors have demonstrated that neutrinos have mass~\cite{KamLAND_rate,bib:prl_rate,bib:reno,dc_bump}, and have searched for the existence of new heavy neutrino states~\cite{bib:Bugey3_osc,bib:neos,danss_osc,stereo_2019,prospect_prd,bib:prl_sterile} and other new physics phenomena~\cite{Hagner:1995bn,TEXONO:2010tnr,DoubleChooz:2012eiq,DayaBay:2018fsh,TEXONO:2018nir,connie_bsm,prospect_dm,Almazan:2021fvo}.  
By observing discrepancies with respect to existing theoretical \nuebar flux and energy spectrum predictions, they have also highlighted limitations of and/or inaccuracies in community fission yield and beta decay databases~\cite{bib:prl_rate,bib:reno_shape,dc_bump,bib:prl_evol,prospect_spec,stereo_shape,DayaBay:2022eyy,bib:fallot,sonzongi2,PhysRevC.98.041303,bib:fallot2}.  
Antineutrino monitoring case studies have explored a variety of potential use case scenarios, such as thermal power load-following and determination of reactor fissile inventory~\cite{hubermon1,hubermon2,bernmon2,huber_mox,Stewart:2019rtd}, and existing \nuebar detectors have confirmed the feasibility of some of these activities~\cite{rovnomon,bowdenmon,bib:prl_evol}.  

The average number of antineutrinos released or detected per nuclear fission depends on the fission isotope in question: different fission isotopes have different fission product yields, with each product varying in its distance from the line of stability and having its own unique nuclear structure and decay scheme.  
Thus, reactor cores with differing fuel compositions are expected to differ in their rate of \nuebar output.  
These expected differences have been explicitly demonstrated in recent \nuebar experiments using the inverse beta decay (IBD) interaction process, $p$ + \nuebar $\rightarrow e^+$+\,$n$, which has a 1.8\,MeV interaction threshold and a precisely-predicted cross-section, $\sigma_{\rm{IBD}}($E$_{\nu})$, versus \nuebar energy E$_{\nu}$~\cite{Vogel:1999zy}.   
For these experiments, measured \nuebar fluxes have been expressed in terms of an IBD yield per fission $\sigma_f$~\cite{reactor_review}: 
\begin{equation}
\sigma_f(t) = \sum_i F_i(t) \sigma_i.
\label{eq:yield}
\end{equation}
In this expression, $F_i(t)$ is the fraction of fissions contributed by isotope $i$ in the sampled reactor core(s) during the experiment's measurement period and $\sigma_i$ its IBD yield per fission,
\begin{equation}
\sigma_i = \int S_i(E_{\nu})\sigma_{\rm{IBD}}(E_{\nu}) dE_{\nu}.
\label{eq:yield2}
\end{equation}
Here, $S_i($E$_{\nu})$ is the true produced \nuebar energy (E$_{\nu}$) spectrum per fission for isotope $i$, and $\sigma_{\rm{IBD}}$ is the inverse beta decay interaction cross-section.   

In a straightforward demonstration of variations in \nuebar emission between fission isotopes, reported IBD yields $\sigma_f$ are clearly offset~\cite{huber_berryman} between measurements at \uFive-burning highly enriched reactor cores~\cite{bib:Krasno1,bib:Krasno2,bib:Krasno3,bib:ILL_nu,bib:nucifer,bib:srp,stereo_rate} and measurements performed at commercial cores burning a mixture of \uFive, \uEight, \pNine, and \pOne~\cite{bib:gosgen,bib:Rovno1,bib:Rovno2,bib:B4,bib:Bugey3,bib:chooz,bib:prl_rate,reno_evol,DoubleChooz:2019qbj}.  
In a separate demonstration, the Daya Bay and RENO experiments have compared IBD yields measured in the same detectors at differing points in their sampled commercial reactors' fuel cycles, observing higher yields during periods with higher (lower) \uFive~(\pNine) fission fractions~\cite{bib:prl_evol,reno_evol}.  

By performing fits to a set of $\sigma_f$ measurements at reactors of well-known fission fraction $F_{i}$, one can use Equations~\ref{eq:yield} and~\ref{eq:yield2} to directly determine the isotopic IBD yield $\sigma_i$ of one or more fission isotopes.  
With a single HEU-based experiment, the IBD yield for \uFive, $\sigma_{235}$, can be trivially determined as $\sigma_{235}= \sigma_f$, since $F_{235}$ approaches unity for these cores.  
On their own, HEU-based $\sigma_{235}$ measurements exhibit deficits~\cite{Giunti} with respect to commonly-used beta-conversion predictions~\cite{bib:huber,bib:mueller2011}, indicating issues in modeling either the core's \nuebar emissions or \nuebar behavior during propagation~\cite{bib:mention2011}.  
Daya Bay and RENO $\sigma_f$ measurements, which encompass multiple data points with differing LEU fuel composition $F_{235,~238,~ 239,~241}$, when combined with modest theoretical constraints on $\sigma_{238,~241}$, yields from the sub-dominant isotopes~\uEight~and~\pOne, enable determination of isotopic yields for both \uFive~and~\pNine~\cite{bib:prl_evol,reno_evol}. These measurements show a deficit with respect to \uFive~conversion predictions, but no such deficit for \pNine, providing further credence to the \nuebar emission mis-modelling hypothesis.  
Going further, global fits of both LEU and HEU datasets can be used to simultaneously determine $\sigma_{235, ~238,~239}$~\cite{surukuchi_flux}: the measured
$\sigma_{238}$~shows a significant (33$\pm$14)\% deficit~\cite{giunti_evol} with respect to summation predictions based on community-standard nuclear databases~\cite{bib:mueller2011}, suggesting potential issues in current \uEight~fission yield measurements or evaluations.  
Future direct determination of isotopic IBD yields for a wider array of fission isotopes beyond \uFive, \pNine, and \uEight, as well as improved precision for these three isotopes, can lead to further understanding or improvement of existing nuclear data, reactor \nuebar models, and reactor-based fundamental physics studies.  

Improved isotopic IBD yield measurements also hold potential benefits for future \nuebar-based applications.  
Some advanced reactor technologies present unique safeguards challenges that may be satisfied by near-field \nuebar-based monitoring capabilities~\cite{Akindele:2021sbh}.  
However, neutrino emissions have never been measured at advanced reactor cores, some of which differ substantially from measured HEU and LEU reactor types in both fuel composition and core neutronics~\cite{doi:10.1080/00295639.2022.2035183,bernmon2,neutrons_shoulder}.  
For example, mixed-oxide reactor fuels, which, unlike conventional low-enriched fuel, are produced from a mixture of uranium and plutonium isotopes, may be deployed in future reactors to realize a closed nuclear fuel cycle or as a means of disposing of existing plutonium stockpiles.  
Fast fission reactor technologies, which, unlike conventional thermal reactors, rely on fast neutron induced fission to maintain criticality, may offer safety and sustainability advantages with respect to conventional reactor types.  
For these reactors, better direct determinations of true underlying $\sigma_i$ can enable more robust and reliable future monitoring capabilities than would be possible using existing demonstrably imperfect models of \nuebar production per fission.  

In this paper, we study how existing and potential future IBD measurements can provide first direct glimpses at \nuebar production by previously unexplored fission isotopes and improve our precision in understanding of the more-studied isotopes \uFive, \pNine, and \uEight.  
By performing loosely constrained fits of isotopic IBD yields to existing LEU and HEU datasets, we demonstrate the feasibility of achieving non-trivial future bounds on \nuebar production by \pOne.  
By applying the same fit techniques to hypothetical future high-precision IBD yield measurements at HEU, LEU, MOX, and fast fission reactors, we show that direct IBD yield determinations for all four primary fission isotopes (\uFive, \uEight, \pNine, and \pOne) can meet or exceed the claimed precision of existing conversion-based predictions while also placing the first meaningful bounds on \pZero~\nuebar production.  

We begin in Section~\ref{sec:global} with a description of the global fit and existing and hypothetical future IBD yield datasets.  
Results of the fit to existing datasets and studies of \pOne~limits are presented and discussed in Sections~\ref{sec:current}.  
In Section~\ref{sec:future}, we describe the set of considered future hypothetical experiments and the result of applying global fits to the hypothetical results of these experiments.  
Main results are then summarized in Section~\ref{sec:summary}.  

\section{Global Datasets and Fit Technique}
\label{sec:global}

In this analysis we perform fits to a set of IBD rate measurements with varying degrees of systematic correlation between each measurement set.  
For an individual measurement, the number of IBD interactions $N$ detected per time interval $t$ can be described as:
\begin{equation}\label{eq:equ_ibd_rate}
N =  \frac{ N_p  \varepsilon
\!P(L)}{4\pi L^2} \int \frac{W_{\textrm{th}}(t) \sigma_f(t)}{\bar{E}(t)}\,dt,
\end{equation}
where
$N_p$ is the number of target protons,
$\varepsilon$ is the efficiency of detecting IBDs, 
$P(L)$ is the survival probability due to neutrino
oscillations, and
$L$ is the core-detector distance.  
Of the time-dependent quantities, $W_{\mathrm{th}}(t)$ is the reactor's thermal power, 
$\bar{E}(t) = \sum_{i}F_{i}(t)e_{i}$ is the core's average energy released per fission,  
$e_i$ is the average energy released per fission of isotope $i$, and $F_i(t)$ and 
 $\sigma_f(t)$, as in Equations~\ref{eq:yield} and~\ref{eq:yield2}, are the fission yields and IBD yields of isotope $i$.  
In order to perform one or multiple measurements of $\sigma_f$, a reactor \nuebar flux experiment must measure $N$ while characterizing the other reactor and detector inputs in Equation~\ref{eq:equ_ibd_rate}.  

\subsection{Existing Datasets}
\label{subsec:exist}

Many experiments have successfully measured $\sigma_f$ values and associated statistical and systematic uncertainties.  
As input for this study, we include time-integrated IBD yield measurements and uncertainties reported by the Goesgen, Bugey-3, Bugey-4, Rovno, Palo Verde, CHOOZ, and Double Chooz LEU-based experiments and the ILL, Savannah River, Krasnoyarsk, Nucifer and STEREO HEU-based experiments, as well as the highly-correlated datasets at varying $F_i$ from the Daya Bay and RENO experiments.  
Calculated fission fractions and measured yields for these experiments, as well as associated uncertainties and cross-measurement systematic correlations, have been summarized in Ref.~\cite{GiuntiGlobal}, and are used for portions of this paper's analysis.  
Input data tables are provided in the public GitHub repository~\footnote{\url{https://github.com/iit-neutrino/AGRIYA}} provided by the authors as an accompaniment to this analysis.  
Since we do not consider short-baseline oscillations as part of this analysis, reactor-detector baselines are not used in analysis of existing datasets, but are nonetheless provided in these tables.  

\subsection{Hypothetical Future Datasets}
\label{subsec:future}

For this study, we also generate hypothetical future IBD yield datasets and uncertainty budgets matching the expected capabilities of experimental deployments at HEU, LEU, MOX and fast reactor types.  
These are also provided in the GitHub supplementary materials, along with assumed uncertainty covariance matrices for all considered hypothetical measurements.  
Hypothetical IBD yield measurements are Asimov datasets free of statistical and systematic fluctuations that are generated according to Equation~\ref{eq:equ_ibd_rate}.  
As input to this equation, fission fractions are required for each host reactor and are described below.   
To match general indications from recent summations~\cite{bib:fallot2} and fission beta~\cite{kopeikin2021}, and \nuebar flux evolution~\cite{bib:prl_evol} measurements, and matching the approach in Ref.~\cite{surukuchi_flux}, we choose input `true' IBD yield values matching a scenario where Huber-Mueller modelled yields~\cite{AnomalyWhite} are only incorrect for \uFive: ($\sigma_{235},~\sigma_{238},~\sigma_{239},~\sigma_{241}$) = (6.05,10.10,4.40,6.03)\,\xSec.  
The yield for \pZero~has not been predicted in the literature to our knowledge, so we estimate it by applying a 3Z-A scaling suggested in Ref.~\cite{sonzogni_insights} to the four previously-mentioned isotopes; the determined central value is $\sigma_{240}$ = 4.96\,\xSec.  
Other experimental assumptions regarding detector, reactor, and experimental layout parameters are then required to define the statistical and systematic uncertainties associated with each hypothetical IBD yield measurement.  

\begin{table*}[tb!]
\centering
\begin{tabular}{l|c|c|c|c|c}
\hline
Parameter & HEU & LEU & MOX & Fast (PFBR) & Fast (VTR) \\ \hline \hline
\textbf{Reactor} &&&&& \\ 
 Thermal Power (MW$_{\textrm{th}}$) & 85 & 2900 & 3200 & 1250 & 300 \\
 Burnup Profile & - & \cite{bib:cpc_reactor} & \cite{Bernstein:2016ayp} & \cite{Behera:2020qwf} & \cite{aap2018} \\ 
 Reactor Cycle Length & 24\,d & 1.5\,y & 1.5\,y & 1.5\,y & 100\,d \\ \hline
\textbf{Experimental} &&&&& \\ 
Core-Detector Distance (m) & 7\,m & 20\,m & 20\,m & 20\,m & 20\,m \\ 
Data-Taking Length & 3\,y & 1.5\,y & 1.5\,y & 1\,y & 100\,d \\ \hline
\textbf{Detector} &\multicolumn{5}{|c}{} \\ 
Active Mass & \multicolumn{5}{|c}{4\,ton (1\,ton)} \\
Target Protons & \multicolumn{5}{|c}{2$\times 10^{29}$ (0.5$\times 10^{29}$)} \\
IBD Detection Efficiency & \multicolumn{5}{|c}{40\%} \\ \hline \hline
\textbf{Uncertainty, Reactor} &&&&& \\ 
Thermal Power & 1.0\% & 0.5\% & 0.5\% & 1.0\% & 1.0\% \\
Fission Fractions & - & 0.6\% & 0.6\% & 0.6\% & 0.6\% \\
Energy per Fission & 0.1\% & 0.2\% & 0.2\% & 0.2\% & 0.2\% \\ \hline
\textbf{Uncertainty, Detector} & \multicolumn{4}{|c}{} \\ 
Target Protons & \multicolumn{5}{|c}{1.0\%} \\
Detection Efficiency & \multicolumn{5}{|c}{0.75\%} \\
IBD Cross Section & \multicolumn{5}{|c}{0.1\%} \\ \hline
\textbf{Total Reactor Systematic} & \textbf{0.5\%} & \textbf{0.8\%} & \textbf{0.8\%} & \textbf{1.2\%} & \textbf{1.2\%} \\ \hline
\textbf{Total Detector Systematic} & \multicolumn{5}{|c}{1.3\%} \\ \hline

\end{tabular}
\caption{Assumed reactor and site parameters for the hypothetical future short-baseline reactor experiments described in the text.}
\label{tab:SBLParams}
\end{table*}

The HEU-based measurement is modeled after the HFIR facility at Oak Ridge National Laboratory, sporting 85\,MW of thermal power, a 100\% \uFive~fission fraction, and a 7\,m reactor-detector center-to-center distance.  
LEU-based measurements are assumed to occur at a 20\,m center-to-center distance from a core following the attributes of a 2.9\,GW$_{\textrm{th}}$ Daya Bay core with an 18\,month fuel cycle.  
Assumed fission fractions are chosen to fall roughly in the middle of the range reported for Daya Bay's cores in Fig.~1 of Ref.~\cite{bib:prl_235239}, and correspond to a fully-loaded core with roughly 1/3 of its rods containing fresh (pure uranium oxide at start-up) fuel; this level of partial reloading is customary when operating cores of this type.  

MOX-based measurements are modelled after the MOX reactor studies of Ref.~\cite{Bernstein:2016ayp}, and is assumed to occur 20\,m from a core with a 3.2\,GW$_{\textrm{th}}$ thermal power and 18\,month cycle length, and fission fractions matching those of the simulated 50\% weapons-grade MOX-burning core.  Weapons-grade (WG) plutonium is characterized by low \pZero~and \pOne~isotopic fractions, and thus a low $F_{241}$ fission fraction at reactor start-up.  
These WG-MOX core parameters correspond to a realizable operational scenario implemented for the goal of plutonium stockpile disposition in a commercial reactor core.  
We will also reference a similar case where 50\% reactor-grade (RG) MOX fuel is used in the same reactor type; these parameters correspond to an operational scenario for a commercial complex operated as part of a closed nuclear fuel cycle program.  
Following recommendations of the authors of Ref.~\cite{Bernstein:2016ayp}, fission fractions for the WG-MOX core example are assumed to match the reported fission fractions for the first third of pictured 50\% WG-MOX running in Ref.~\cite{Bernstein:2016ayp}, while fractions from the RG-MOX case are assumed to match the those of 50\% WG-MOX running between days 800 and 1350~\cite{Anna}; fission fractions were extracted by interpolating fission rates from this reference and normalizing such that the sum for the four primary fission isotopes is equal to unity.  

It should be stressed that modeled fuel content evolution for LEU and MOX cores is highly dependent on the initial conditions of the fuel, on the neutronics of the involved core type, and on reactor operations.  
In this study, we include one specific fission fraction set for each fuel type -- LEU, WG-MOX, and RG-MOX; the impact or potential benefits of further variations between LEU or MOX core types is not considered.  

\begin{figure}[phtb!]
  \centering
  \includegraphics[trim={0 0 0 1cm},clip, width=0.5\textwidth]{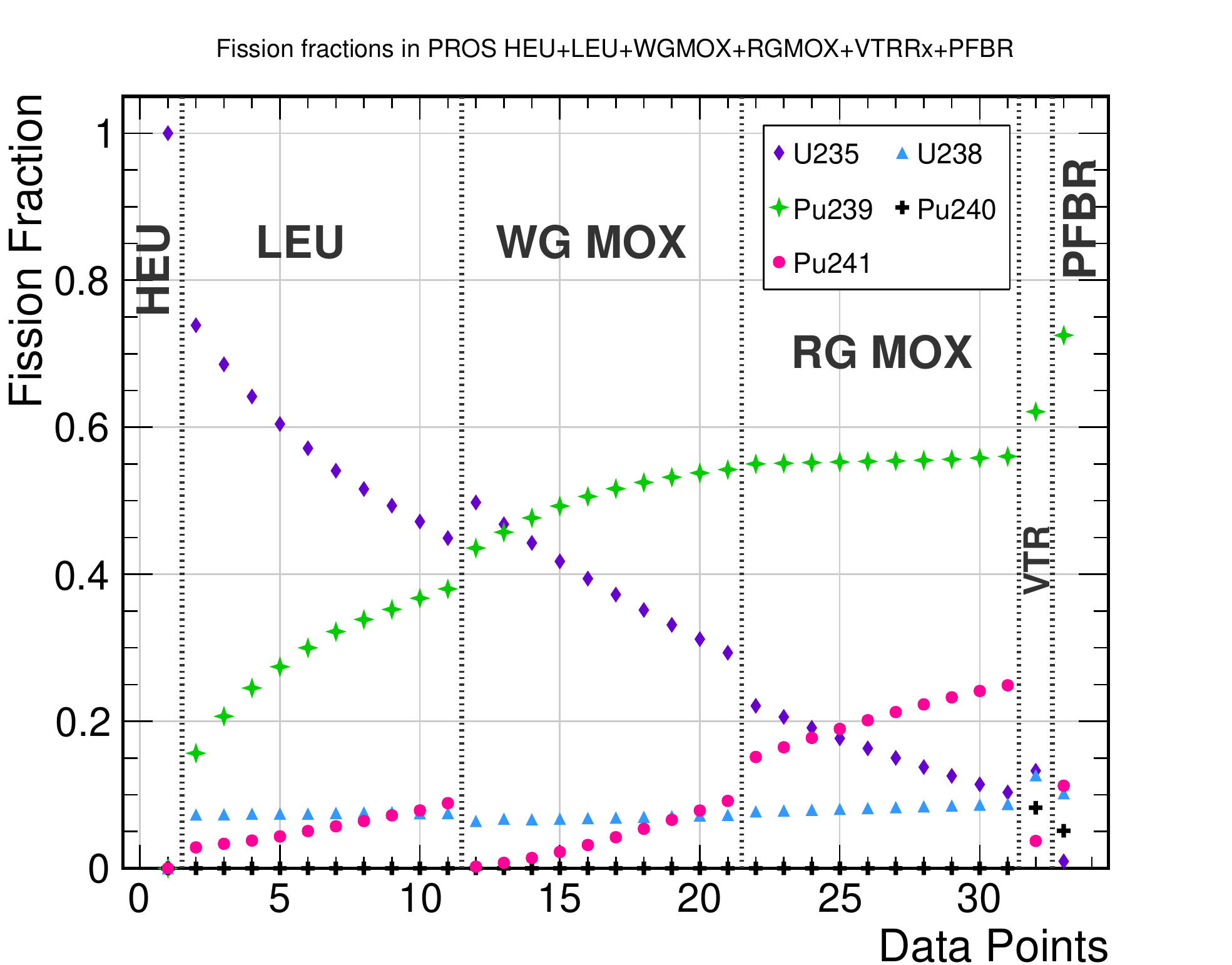}
  \caption{Fission fractions used for hypothetical future measurement data points described in this section. See text for details.}
  \label{fig:FissionFrac}
\end{figure}

Finally, two experiments are assumed to occur at the baselines of 20\,m and 7\,m distances from the primarily plutonium-burning 1.25 GW$_{\textrm{th}}$ PFBR fast breeder reactor in India~\cite{Behera:2020qwf} and the 300 GW$_{\textrm{th}}$ Versatile Test Reactor fast reactor~\cite{aap2018} respectively.
The former reactor plays a central role in plans for realization of an independent, sustainable nuclear fuel cycle in India, while the latter has been developed as a US-based reactor materials and irradiation R\&D facility based at Idaho National Laboratory~\cite{doi:10.1080/00295639.2022.2035183}.  
Assumed reactor and site parameters for all measurements are summarized in Table~\ref{tab:SBLParams}; fission fraction values for all hypothetical measurement data points used in this study are illustrated in Figure~\ref{fig:FissionFrac}.  

For all experiments, an IBD detector matching qualities of the 4\,ton PROSPECT reactor \nuebar detector are used~\cite{prospect_nim}; relevant parameters are also listed in Table~\ref{tab:SBLParams}.  
In some cases, a 1~ton detector with otherwise similar experimental parameters is also considered; this case enables investigation of the value of using a near-future compact \nuebar monitoring detector, such as the Mobile Antineutrino Demonstrator~\cite{Bowden:2022rjt}~(MAD), to perform IBD yield benchmarking measurements at multiple reactor locations.  
In all cases, the statistical uncertainties associated with each datapoint for each reactor-detector combination are estimated using the associated detector and reactor parameters quoted in Table.~\ref{tab:SBLParams} and lie between 0.15 \% and 0.2 \%.  
For simplicity, we do not consider statistical and systematic uncertainty contributions from IBD-like backgrounds; for a PROSPECT-like detector expecting signal-to-background ratios of better than 4 (10) deployed on-surface at an HEU (LEU) reactor~\cite{Andriamirado:2021qjc}, IBD counts would be expected to dominate measurement statistical uncertainties.

Hypothetical measurements should also be accompanied by predicted reactor- and detector-related systematic uncertainties, which are also summarized in Table~\ref{tab:SBLParams}.  
Systematics for most cores are dominated by the uncertainty in the thermal power produced by the operating core.  
Commercial reactor companies have provided sub-percent precision in reported thermal powers for existing IBD yield measurements at numerous reactor sites~\cite{bib:caojPower,conant_thesis}; for this reason, we choose 0.5\% uncertainty for LEU and MOX cores.  
While similar thermal power measurement devices and strategies could be applied to HEU facilities, in practice, legacy systems used in existing HEU facilities have recently provided thermal power uncertainties closer to 2\%~\cite{stereo_rate}; for this analysis, we optimistically assume implementation of upgraded measurement systems or techniques capable of providing 1\% precision at an HEU core.  
Advanced technologies for time-stable and high-precision thermal power monitoring in sodium-cooled fast reactors like PFBR and VTR are under active development, due to the difficulties associated with the coolant's high temperature and chemical corrosiveness; given the lack of available quantitatively demonstrated capabilities, we also assume a 1\% thermal power uncertainty for this core type.  While thermal power uncertainties for different reactors are assumed to be uncorrelated, this uncertainty is correlated between multiple measurements at the same core.  
Measured IBD yields for an experiment will also be uncertain due to the limits in knowledge of fission fractions in the core, which is defined via detailed reactor core simulations.  
In the absence of these calculations for all core types, we will assume an uncertianty of 0\% for the HEU experiment and 0.6\% for all other cores, following the value quoted by Daya Bay and others for LEU cores~\cite{bib:cpc_reactor}.  
This uncertainty is also assumed to be uncorrelated between cores, but correlated between measurements at the same core.  
Isotopic energy release per fission $e_i$ -- required for calculating expected experiment statistics -- have minor IBD yield uncertainty contributions of 0.1\% to 0.2\% depending on  core fuel content~\cite{bib:fr_ma}; the  $e_i$ central value and uncertainty for \pZero~is assumed to match that of \pOne.  

On the detector side, uncertainties are dominated by the limited knowledge of IBD detection efficiency, assumed to be known with 0.75\% precision, as well as knowledge of the total number of protons within the detector's target region, assumed to be known to 1\%; these chosen values reflect those achieved in a range of recent large- and small-detector IBD experiments~\cite{stereo_rate, bib:prd_rate,reno_evol,dc_nature}.  
In this analysis, we consider the possibility of moving a single reactor neutrino detector to multiple reactor core types to perform systematically correlated IBD yield measurements; for this reason, unless otherwise mentioned, we treat detector systematic uncertainties as correlated between all measurements.  

\subsection{Global Fit Approach}

To obtain isotopic IBD yields in this analysis, we use a least-squares test statistic:
\begin{equation}
\begin{gathered}
\label{eq:Iso3}
\chi^2 = \sum_{a,b}\bigg(\sigma_{f,a} - r \sum_i F_{i,a} \sigma_i\bigg)
	\textrm{V}^{-1}_{ab}
    \bigg(\sigma_{f,b} - r \sum_i F_{i,b} \sigma_i\bigg) \\
     + \sum_{j,k}(\sigma^{th}_{j}-\sigma_{j}) \textrm{V}^{-1}_{\textrm{ext},jk}(\sigma^{th}_{k}-\sigma_{k}).
\end{gathered}
\end{equation}
In this fit, experimental inputs $F_i$ and $\sigma_f$ are as described above, and the sum $i$ is run over five fission isotopes, \uFive, \uEight, \pNine, \pOne, and \pZero, with five attendant IBD yield fit parameters.    
The experimental covariance matrix V defines the uncertainties for each experiment and their cross-correlations, as described in the previous-subsection.  
The final term is used to constrain fitted $\sigma_i$ values to theoretical predictions by adding a penalty that increases as the two quantities diverge.  
In contrast to most recent global IBD yield fits~\cite{Giunti, Giunti2, bib:prl_evol}, we are interested in examining weakly-constrained or un-constrained simultaneous fits of all relevant fission isotopes' IBD yields.  
For this reason, $j$ and $k$ sum only over the three sub-dominant isotopes,  \uEight, \pOne, and \pZero, and the 3$\times$3 $\textrm{V}^{-1}_{\textrm{ext}}$ is diagonal (no assumed uncertainty correlation between isotopes), with elements set to achieve wide $1\sigma$ theoretical constraints of 75\% of the predicted yield.  
To compare to previous IBD yield fits~\cite{surukuchi_flux,giunti_diagnose}, we occasionally consider the much tighter (2.6\%) bounds on $\sigma_{241}$ quoted by the Huber model~\cite{bib:huber}.  
For fits not involving fast reactor datasets, $\sigma_{240}$ is pegged to the theoretically-predicted value, and has no effect on the subsequent 4-parameter fit.  


\section{Fits to Existing Datasets and \pOne~IBD Yield Constraints}
\label{sec:current}

\begin{figure}[phtb!]
  \centering
  \includegraphics[trim={0.6cm 0 0.5cm 0 },clip,width=0.5\textwidth]{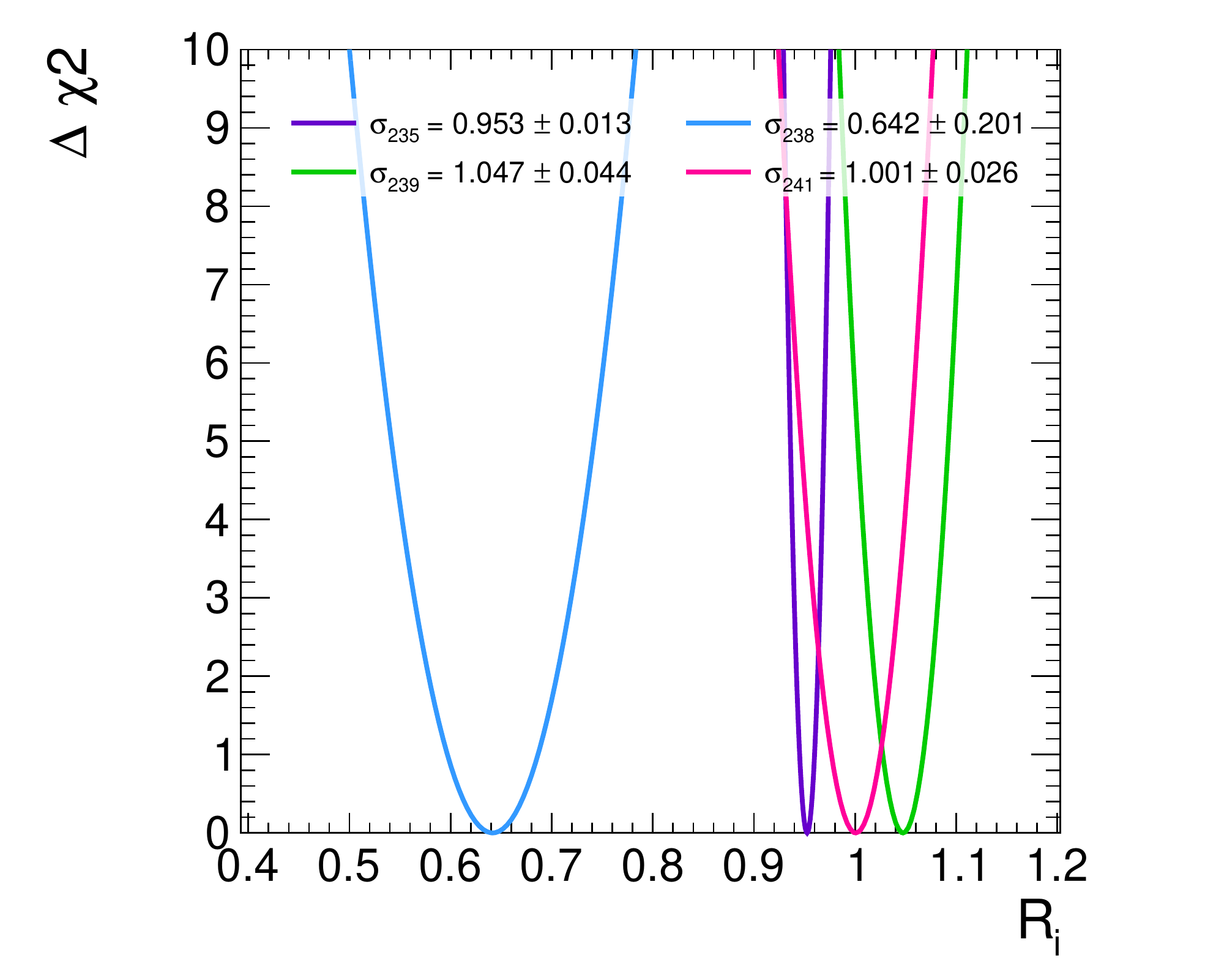}
  \includegraphics[trim={0.6cm 0 0.5cm 0 },clip,width=0.5\textwidth]{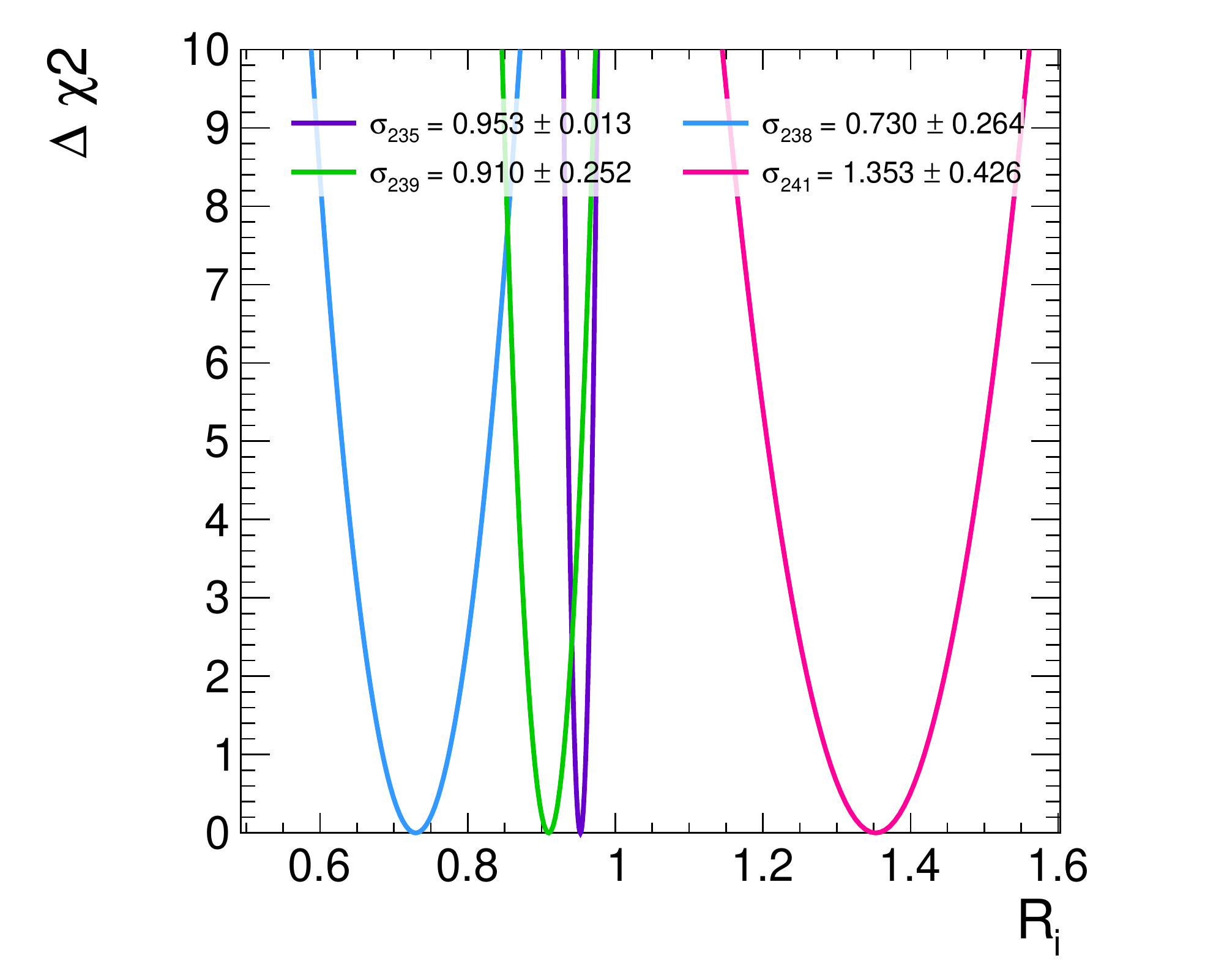}
  \caption{Isotopic IBD yield fit results for the existing global dataset with tight (top, 2.6\%) and loose (bottom, 75\%) external constraints on the \pOne~yield.  Test statistic values with respect to the best-fit ($\Delta\chi^2$) are shown versus input value for each of the four primary fission isotopes.  For each isotope's curve, the fit is marginalized over the other isotopes.}
  \label{fig:chi2}
\end{figure}

We first consider IBD yield fits applied to the existing global yield datasets described briefly in Section~\ref{subsec:exist}.  
By first applying tight 2.6\% constraint on \pOne, we largely reproduce unconstrained \uFive, \uEight, and \pNine~yield best-fit values reported for the oscillation-free fit in Ref.~\cite{giunti_diagnose}.  
Test statistic values with respect to the best-fit ($\Delta\chi^2$) versus input value are shown for each isotope in Figure~\ref{fig:chi2}, while minimizing over the three other isotopic yield parameters.  
We observe a best-fit \uFive~yield more than 3$\sigma$ (5\%) below the Huber-predicted value, and a best-fit \uEight~yield that deviates from the predicted central value by (36$\pm$20), slightly more than in previous fits~\cite{giunti_diagnose}.  
As in previous fits, the \pNine~yield is found to be consistent with Huber-predicted values within a 5\%, $\sim 1 \sigma$ uncertainty band.  
This similarity in results indicates that the relatively new STEREO data point~\cite{stereo_rate}, while qualitatively bolstering confidence in historical observations of a $\sim$5\% yield deficit at HEU cores~\cite{STEREO:2020fvd}, has fairly modest quantitative impact on the primary issues surrounding data-model agreement for conversion-predicted uranium IBD yields.  

With consistency established with respect to previous analyses, we proceed with loosening of yield constraints for all fission isotopes.  
Figure~\ref{fig:chi2} shows reported isotopic $\Delta \chi^2$ test statistic values versus input $\sigma$ value for each isotope while applying a looser constraint on \pOne~of 75\%.  
Best-fit parameters and 1$\sigma$ ranges are found to be: 
\begin{equation}
\begin{gathered}
\label{eq:Iso3}
\sigma_{235} = 6.37 \pm 0.08; \\
\sigma_{238} = 7.37 \pm 1.95;  \\
\sigma_{239} = 4.00 \pm 1.01;  \\
\sigma_{241} = 8.16 \pm 3.47.\\
\end{gathered}
\end{equation}
The best-fit $\chi^2_{min}$ is found to be 26.2 for 38 degrees of freedom (41 data points, 3 fit parameters), indicating an acceptable goodness-of-fit.  
However, this value is only slightly lower than that provided by the more-constrained fit ($\chi^2_{min}$ = 26.6), indicating that this enhanced freedom has not substantially improved data-model agreement.  
Central values of \uFive, \uEight, and \pNine~fit parameters are relatively stable, remaining within 15\% of those provided by the more-constrained fit.  
Meanwhile, the newly freed \pOne~yield increases by 35\%, although $\sigma_{241}$ nonetheless remains consistent with its model-predicted value within its large 43\% relative uncertainty band.  
Thus it appears that the current global IBD yield dataset does not have the statistical power to provide meaningful tests of underlying modelling issues for~\pOne.  
The disappointing lack of new insight should not be too surprising, given the  small ($\mathcal{O}$(5\%) or less) fractional contribution of \pOne~fissions in all existing measured reactor cores.  

\begin{figure*}[tbph!]
  \centering
  \includegraphics[width=1\textwidth]{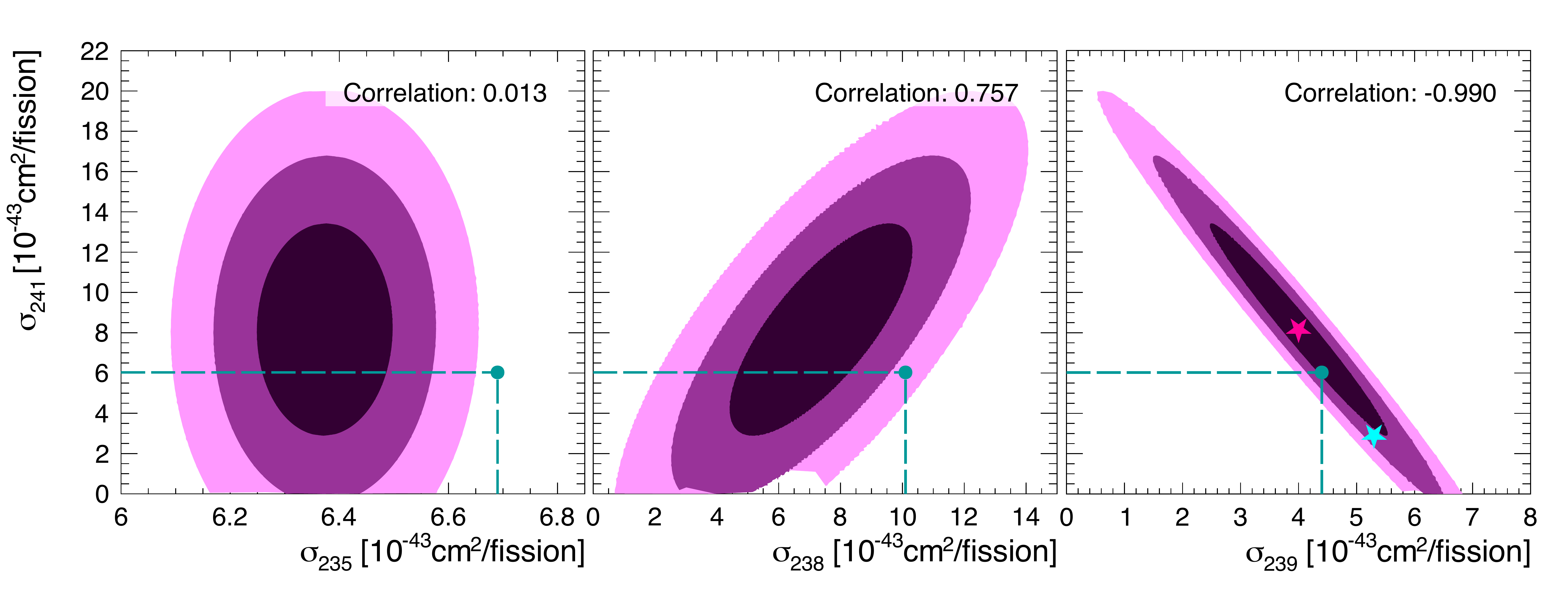}
  \caption{Isotopic IBD yield fits for the existing global dataset with loose (75\%) external constraints on the \pOne~IBD yield, $\sigma_{241}$.  Contours are pictured for $\sigma_{241}$ relative to the other isotopic yields, with the fit marginalized over the non-pictured isotopes.  Correlation coefficients between fitted $\sigma_{241}$ and the other yields are given in the plot legends. Also shown in dashed lines are the theoretical IBD yields predicted by the Huber-Mueller model. Stars indicate IBD yields chosen for illustration in Fig.~\ref{fig:slope}.}
  \label{fig:TrianglePlot}
\end{figure*}

\begin{figure}[phtb!]
  \centering
  \includegraphics[width=0.5\textwidth]{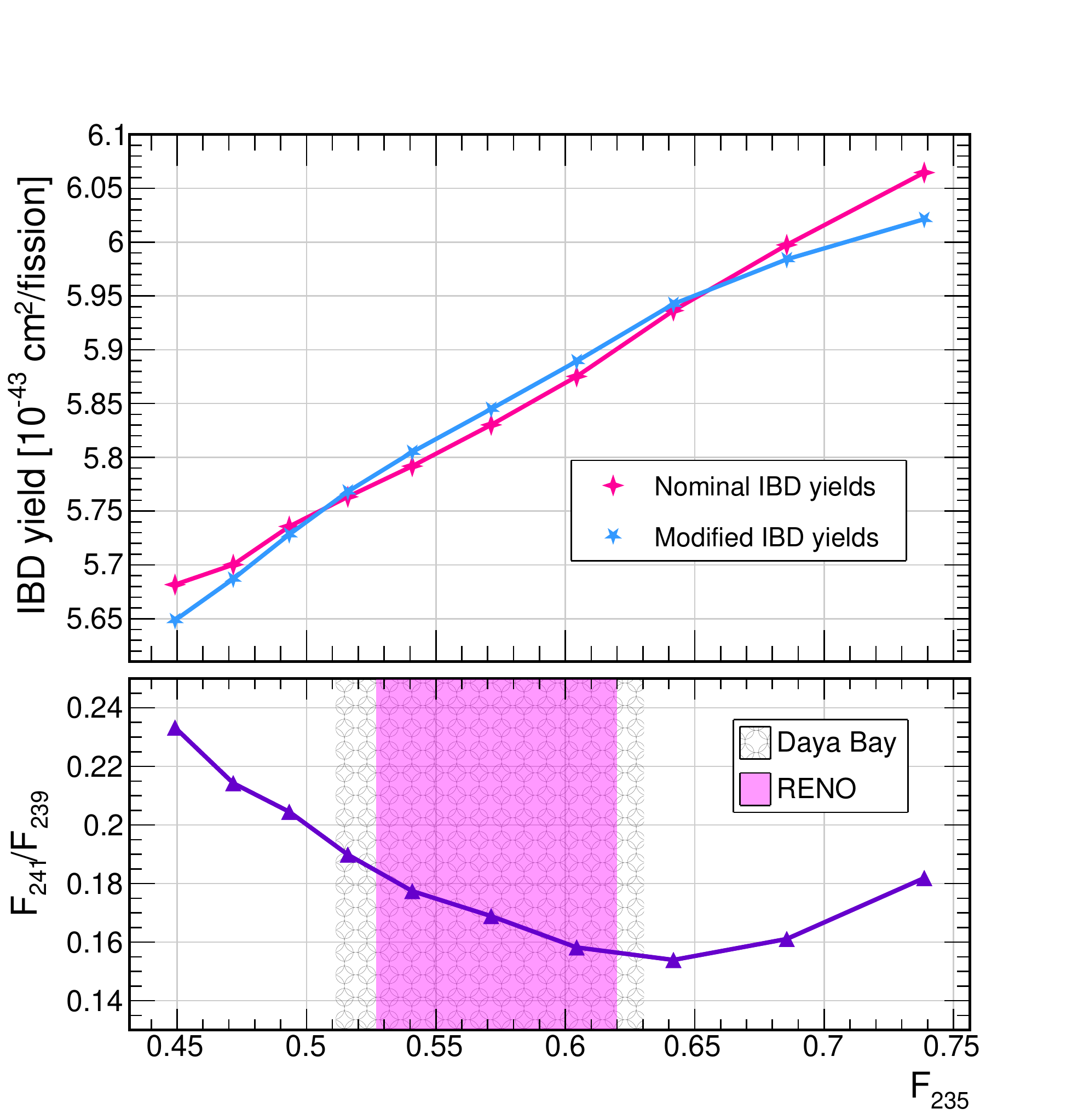}
  \caption{Top: IBD yield sets for two hypothetical LEU measurements: one assuming measurements align with isotopic IBD yields matching the best-fit for the existing global dataset, and another assuming alignment with $\sigma_{239}$ and $\sigma_{241}$ values matching those indicated in Figure~\ref{fig:TrianglePlot}.  The latter scenario's values lie outside of the 1 $\sigma$ region preferred by the global IBD yield dataset; for this scenario, $\sigma_{238}$~is reduced to enable better vertical alignment of the two datasets and easier comparison of slopes.  Bottom: Ratio~(F${241}$/F$_{239}$) of the fission yields of \pOne~and\pNine~for the hypothetical LEU dataset.  Realized F$_{235}$~ranges for RENO and Daya Bay datasets are also pictured.}   
  \label{fig:slope}
\end{figure}
However, it is interesting to note that $\sigma_{241}$ $1\sigma$ error bands are found to be tighter than the externally-applied constraint.  
This indicates that there \emph{are} features in the existing global dataset that provide the power to specifically constrain \pOne.  
To attempt to identify these features, we examined correlations between fitted isotopic yields, which are depicted in Figure~\ref{fig:TrianglePlot} as best-fit parameter space regions in two dimensions between \pOne~ and the other three isotopes.  
Substantial \pOne-\pNine~and \pOne-\uEight~degeneracies can be observed, with the former reflected in a more than five-fold increase in uncertainties on $\sigma_{239}$ between the more-constrained (4\% uncertainty) and less-constrained (25\% uncertainty) fits.  
Degeneracies can also be expressed by calculating correlation coefficients between the fitted yield parameters, which are also given in the legends of Fig~\ref{fig:TrianglePlot}:
\begin{equation}
\label{eq:corr}
\rho_{\sigma_{i},\sigma_{j}}= 
\frac{\overline{ (\sigma_{i} - \overline{\sigma_{i}}) 
(\sigma_{j} - \overline{\sigma_{i}})}}
{\sigma_{\sigma_{i}}\sigma_{\sigma_{j}}}
\end{equation}
The extreme \pOne-\pNine~correlation can be understood by observing the fission fraction evolution trends experienced by LEU reactors, as depicted in Figure~\ref{fig:FissionFrac}.  
In these cores, $F_{239}$ and $F_{241}$ rise in tandem with reactor fuel burn-up, making it hard for unconstrained fits to simultaneously determine both  $\sigma_{239}$ and $\sigma_{241}$.  
It can also be understood as a simple reality of underlying nuclear physics in the core: \pOne~is produced by via multi-neutron capture on \pNine, and thus its build-up in the core is dependent on the build-up of the latter.  
In aspects of previous multi-datapoint LEU analyses, such as those of Daya Bay~\cite{bib:prl_evol,DayaBay:2021dqj} and RENO~\cite{reno_evol}, \pOne~and \pNine~fission fractions are treated as explicitly linearly correlated.

We examine the limits of this linear correlation by generating hypothetical LEU reactor IBD yield datasets following the method described in Section~\ref{subsec:future} and fission yields from Figure~\ref{fig:FissionFrac}: one dataset assumes isotopic yields matching the best-fit for the existing global dataset, and the other assumes true \pOne~and \pNine~yields close to the axis of anti-correlation between the two datasets, but beyond the $1\sigma$ bounds allowed by the data.  
Chosen true yields for this test are illustrated in the right panel of Figure~\ref{fig:TrianglePlot};  the \uEight~ yield for this case, 8.8\,cm$^2$/fission, was chosen to vertically align the two yield datasets for easier comparison of trends.  
Hypothetical yields for these two cases are pictured in Figure~\ref{fig:slope}.  
The test cases clearly differ in the change in slope, or curvature, present in the LEU data points, providing an indication of the primary source of unique \pOne~yield information in current and future experimental data.  
The extreme \pNine-\pOne~yield offset in this example amplifies the impact of a modest non-constant relationship between $F_{239}$ and $F_{241}$ in LEU-based datasets, which is also illustrated in Figure~\ref{fig:slope}.  
To test the validity of this hypothesis with existing datasets, we perform a fit to only the RENO and Daya Bay LEU datasets while applying loose 75\% external constraints on all four isotopes.  
While large uncertainty increases are seen in $\sigma_{235}$ and $\sigma_{238}$, $\sigma_{239}$ and $\sigma_{241}$ fractional uncertainties are altered by $<$30\%, and fractional bounds on $\sigma_{241}$ (43\%) remain tighter than the 75\% external constraint.  
Thus, in the existing global dataset, it does appear that that the Daya Bay and RENO LEU data points are responsible for the modest breaking of degeneracy between \pNine~and \pOne~yields.  

Adding this to previously-established trends, it is straightforward to recount the independent features of the global IBD yield dataset that enable determination of all four isotopes' IBD yields: 
\begin{itemize}
    \item HEU-based experiments' $\sigma_f$ measurements directly constrain $\sigma_{235}~$\cite{Giunti}.  
    \item The measured relative linear $\sigma_f$ slope versus fuel burn-up at LEU-based experiments directly constrains $\sigma_{239}$~\cite{bib:prl_evol}.  
    \item The time-integrated offset in $\sigma_f$ between HEU and LEU cores constrains $\sigma_{238}$~\cite{surukuchi_flux}.  
    \item The curvature of $\sigma_f$ slope versus fuel burn-up at LEU experiments constrains $\sigma_{241}$.  
\end{itemize} 
As we move on to consider possible future IBD yield measurement scenarios, these high-level principles serve to guide attention toward those with particular promise for improving global knowledge of isotopic yields.  
In particular, we will look to explore new multi-dataset measurements that can provide an enhanced view of $\sigma_f$ curvature with host reactor fuel evolution.  

We end this Section by noting that within the current global dataset, Daya Bay contains currently-unexploited potential.  
Ref.~\cite{bib:prl_235239} indicates $\mathcal{O}$(5\%) F$_{241}/$F$_{239}$ variations between reactor cycles that are averaged out in its current fuel content binning scheme.  
To estimate the achievable gains in the fission yields, we generate an Asimov IBD yield dataset with fission fractions taken from a combination of rates, RENO and Daya Bay-like experiment is divided into two halves; one with the default fission fractions while the other having F$_{241}/$F$_{239}$ relatively reduced by 2.5\%.
The systematic and statistical uncertainties are assumed to match the existing global dataset and the yields are generated using best-fit results from the global dataset.
Such a joint fit provides a modest improvement in the precision of fission yield of ($\sigma_{235}, \sigma_{238}, \sigma_{239}, \sigma_{241}$) = (1.3\%, 24.8\%, 19.7\%, 39.2\%) compared to the precision of (1.3\%, 26.4\%, 25.2\%, 42.6\%) for the existing global dataset.
If we further double the statistics of the Daya Bay Asimov data--as expected from the full Daya Bay dataset---in conjunction with the modified binning in fission fractions, we find a further improvement in precision to (1.3\%, 21.7\%, 16.4\%, 30.8\%).
Thus, we conclude that it may be worthwhile for Daya Bay to consider a more diversified fuel content binning scheme in a future analysis of its final full-statistics IBD yield dataset.  
This observation may also be applicable to other high-statistics datasets spanning many LEU reactor cycles, such as those recorded by RENO and DANSS~\cite{danss_osc}.  



\section{Future Improvements From New Measurements at Multiple Core Types}
\label{sec:future}

We now turn to consideration of future improvements in global knowledge of isotopic IBD yields by performing new measurements at a range of different reactor core types.  
We will begin by considering the most imminently-achievable next steps: short baseline measurements of a single LEU core over a full fuel cycle, and a subsequent systematically-correlated measurement at an HEU using the same \nuebar detector.  
We will then proceed to study possible improvements gained by making measurements at mixed-oxide and plutonium-burning fast reactor core types.  

\subsection{Benefits of New HEU and LEU Measurements}

Some benefits of new measurements of IBD yields at short distances from a full LEU reactor core cycle have already been discussed in the literature~\cite{surukuchi_flux}, and have served as part of the physics motivation for the NEOS-II experiment~\cite{neos2}.  
In particular, this configuration enables access to a wider range of $F_{239}$ and $F_{235}$ values beyond those achieved at $\theta_{13}$ experiments sampling multiple cores, which should result in improved $\sigma_{239}$ constraints.  
When coupled with a systematically-correlated HEU-based measurement, which could be achieved via two site deployments of the same detector system, direct constraints on $\sigma_{238}$ may exceed the claimed precision of the summation prediction of Mueller \emph{et al.}~\cite{bib:mueller2011}.  
Multiple current or near-future efforts, such as  PROSPECT-II~\cite{Andriamirado:2021qjc} or MAD~\cite{Bowden:2022rjt}, are well-suited to realize part or all of this combined LEU-HEU measurement program.  

Such a setup would also broaden access to LEU fuel content regimes with less linear relationships between $F_{239}$ and $F_{241}$, allowing for improved constraint of $\sigma_{241}$.  
This improvement was demonstrated above for the hypothetical LEU measurements in Figure~\ref{fig:slope}.  
Realized effective $F_{239}$ ranges for Daya Bay and RENO are also highlighted with shaded bands; we note that offsets in median F$_{235}$ (and, while not pictured, also F$_{241}$/F$_{239}$) between hypothetical LEU and Daya Bay/RENO cases is due to the specifics of the single cycle core loading simulated in Ref.~\cite{bib:cpc_reactor}.  
A new short-baseline LEU measurement set can capture periods earlier and later in the fuel cycle of a conventional LEU core with respect to RENO and Daya Bay, when relative contributions of \pNine~and ~\pOne~fissions deviate most strongly from their cycle-integrated mean.  
For the hypothetical short-baseline LEU measurement, F$_{239}$/F$_{241}$ varies roughly 6\%, from 17\%  to 23\%, over a cycle.  
Daya Bay's and RENO's $F_{241}$/$F_{239}$ ratios, meanwhile vary by only 3\% or less, with maximums and minimums of 20\% and 17\%, respectively~\cite{bib:prl_evol,reno_evol}.  

\begin{table*}[thbp!]
\centering
\begin{tabular}{c|c||c|c|c|c|c}
\hline
\multirow{2}{*}{Case} & \multirow{2}{*}{Description} & \multicolumn{5}{|c}{Precision on $\sigma_i$ (\%)} \\ \cline{3-7}
& & \uFive &\uEight & \pNine & \pZero & \pOne \\ \hline \hline
- & Existing Global Data &  1.3 & 26.4  & 25.2 & - & 42.6 \\  \hline \hline 
1& HEU + LEU & 1.6 & 11.1 & 4.6 & - & 10.5 \\ \hline
3& HEU + LEU + RG-MOX & 1.6 & 9.7 & 2.2 & -& 3.4\\ \hline
2& HEU + LEU + WG-MOX & 1.6 & 9.9 & 2.5 & -& 3.6\\ \hline
4& HEU + LEU + Fast & 1.6 & 10.9 & 4.6 & 27.2 & 10.3 \\ \hline
\bf{5} & \bf{All} & \bf{1.6} & \bf{9.5} & \bf{2.1} & \bf{23.6} & \bf{3.3} \\ \hline
6& All, Uncorrelated  & 1.5 & 14.3 & 2.1 & 36.2 & 4.2 \\ \hline\hline
- & Model Uncertainty~\cite{GiuntiGlobal} & 2.1 & 8.2 & 2.5  & - & 2.2 \\ \hline
\end{tabular}
\caption{Constraints on IBD yields of \uFive, \uEight, \pNine, \pZero, and \pOne,~from future hypothetical datasets from LEU and HEU reactors, given as a percentage of the best fit yield.  
For all cases unless noted, detector systematic uncertainties are assumed to be correlated between measurements, and a 75\% external constraint is used for \pOne~and for \pZero~when applicable.  The `All' case considers inclusion of HEU, LEU, RG-MOX, VTR  and PFBR yield measurements employing the same detector.   
Model prediction uncertainties from ~\cite{GiuntiGlobal} are also provided.  
}
\label{tab:Uncertainty_Improvement}
\end{table*}

The extent to which these HEU and LEU measurements can improve constraints on $\sigma_{241}$ has so far not been investigated in the literature.  
To do so, we apply the four-parameter yield fit of Eq.~\ref{eq:Iso3} to the hypothetical HEU and LEU datasets described in Section~\ref{subsec:future},  Table~\ref{tab:SBLParams}, and Figure~\ref{fig:FissionFrac}.  
Table~\ref{tab:Uncertainty_Improvement} gives the resulting precision in measurements of the four isotopic IBD yields probed by this new HEU+LEU dataset.  
The most striking difference with respect to the current global dataset is the substantial improvement in knowledge of \pNine~and \pOne~yields.  
Uncertainties in $\sigma_{239}$ and $\sigma_{241}$ are improved from 25.2\% and 42.6\% in the existing dataset to 4.6\% and 10.5\%, respectively, greater than four-fold improvement in both values.  
As illustrated in Figure~\ref{fig:TriangleNew}, this improvement can be partially attributed to the reduction in degeneracy between these two isotopes' fission fraction variations over a full LEU fuel cycle.  
If all measurements are instead performed with a 1\,ton detector, more closely approximating the expected size of the MAD detector, uncertainties are similar in size, with $\sigma_{235, 238, 239, 241}$ shifting from (1.6\%, 11.2\%, 4.6\%, 10.5\%) for the PROSPECT-II sized detector case to (1.62\%, 11.7\%, 6.1\%, 14.6\%) for the MAD detector case.  
Thus, the HEU+LEU deployment scenario may yield major benefits for both physics-oriented or smaller applications-oriented future detectors.  

As noted in Ref.~\cite{surukuchi_flux}, $\sigma_{238}$ constraints are also significantly improved, primarily due to the correlated nature of the detector systematics assumed between the HEU and LEU measurements.  
If this correlation is removed, or if the chosen optimistic 1\% HEU thermal power uncertainties are increased to the currently-achievable 2\% level, precision in knowledge of the \uEight~yield is substantially reduced -- to 18.1\% and 17.2\% for these two cases, respectively -- while precision in knowledge of the \pOne~yield is virtually unchanged.  
Thus, following the next generation of short-baseline HEU and LEU measurements, the precision of knowledge of the \pOne~yield may rival that of its sub-dominant \uEight~counterpart, and will be less dependent on a detailed understanding of host reactors' thermal powers and on movement-induced changes in detector response.  
At this point, direct \nuebar-based measurements of \pOne~fission attributes may begin to have useful application in testing the general accuracy of nuclear data knowledge for this isotope -- similar to the value provided by \nuebar-based constraints of \uEight~from the current global dataset.  

\subsection{Benefits from MOX Reactor Measurements}

\begin{figure*}[btph!]
  \centering
  \includegraphics[trim={0.6cm 0 0.5cm 0 },clip,width=0.33\textwidth]{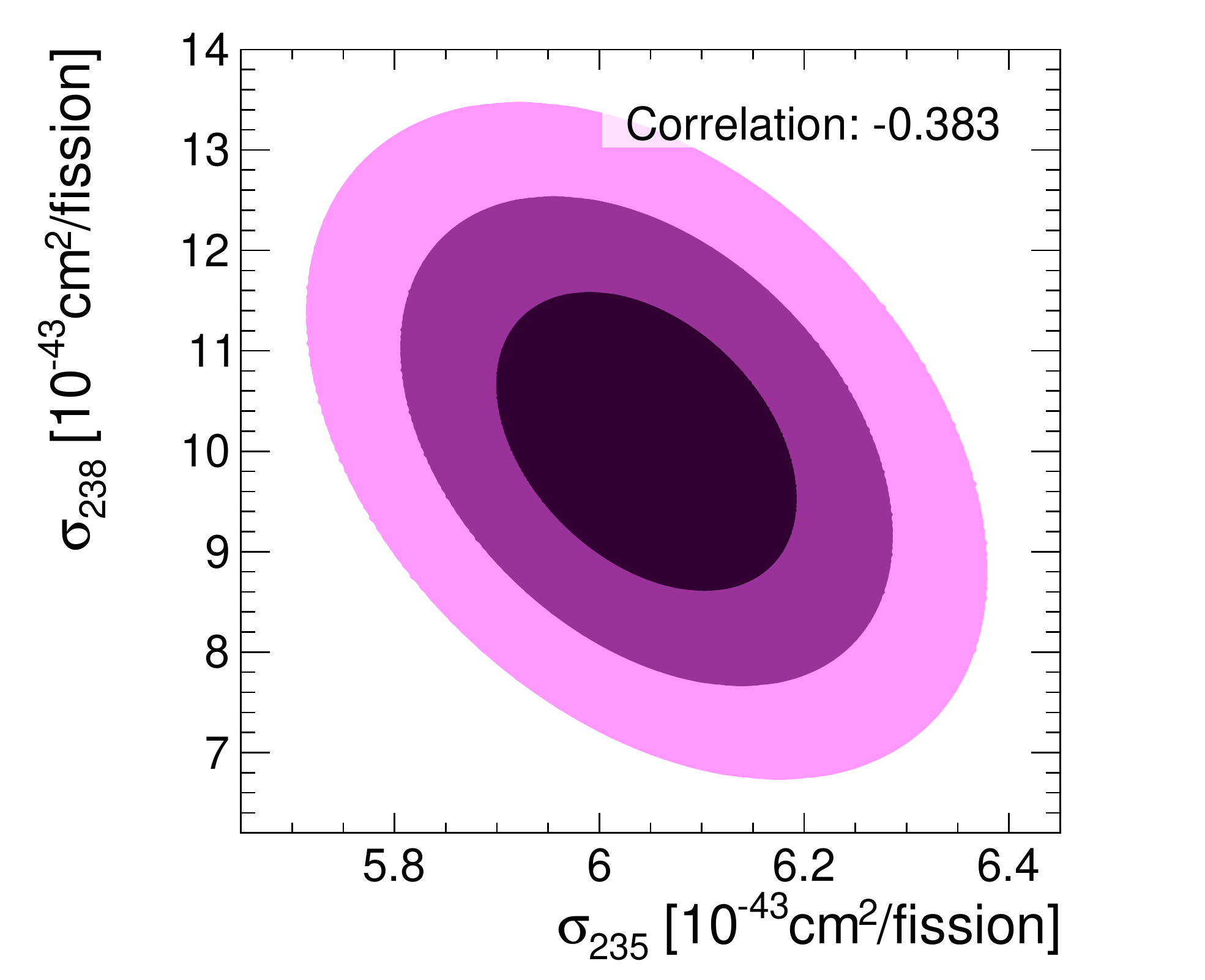}
  \includegraphics[trim={0.6cm 0 0.5cm 0},clip,width=0.33\textwidth]{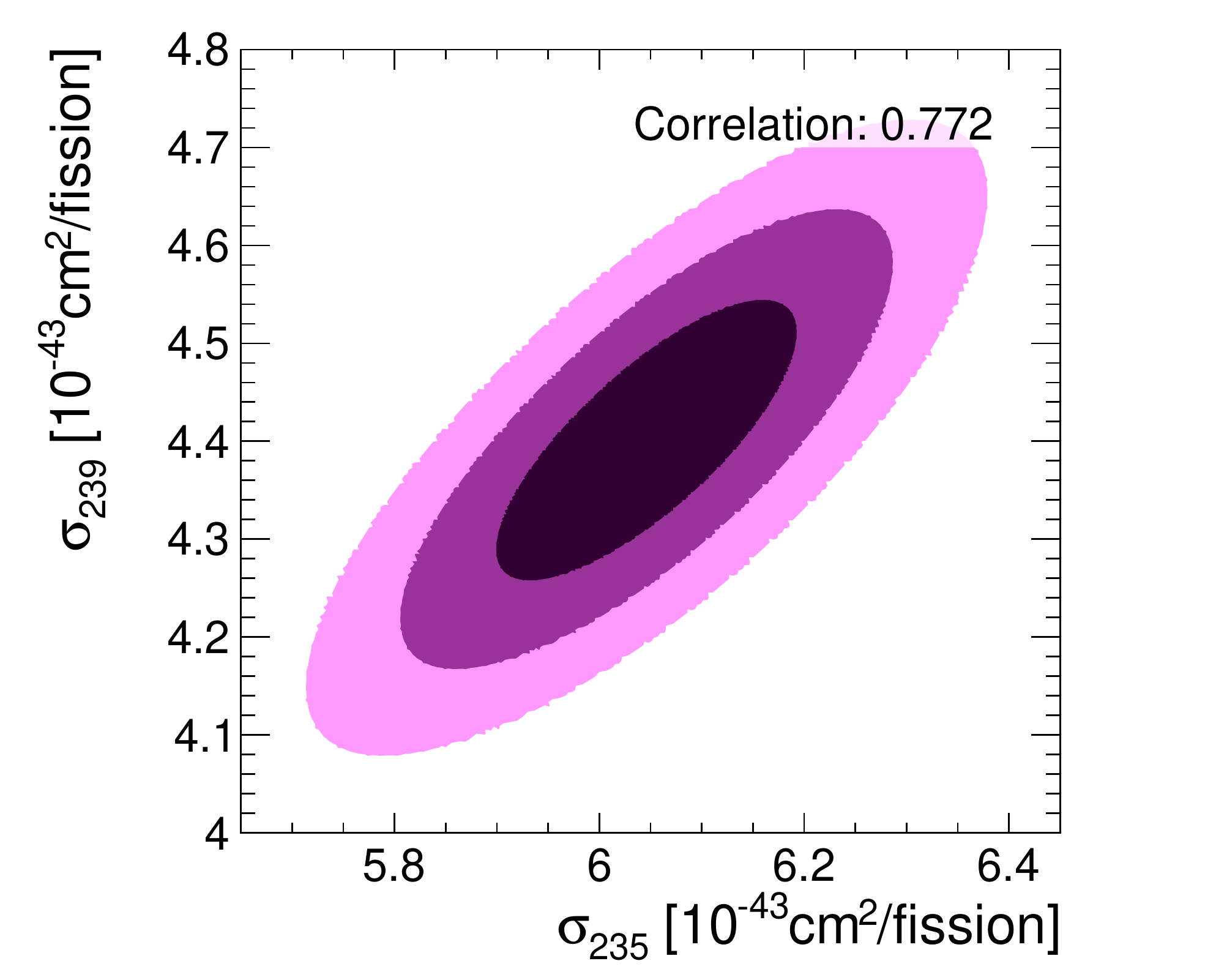}
  \includegraphics[trim={0.6cm 0 0.5cm 0},clip,width=0.33\textwidth]{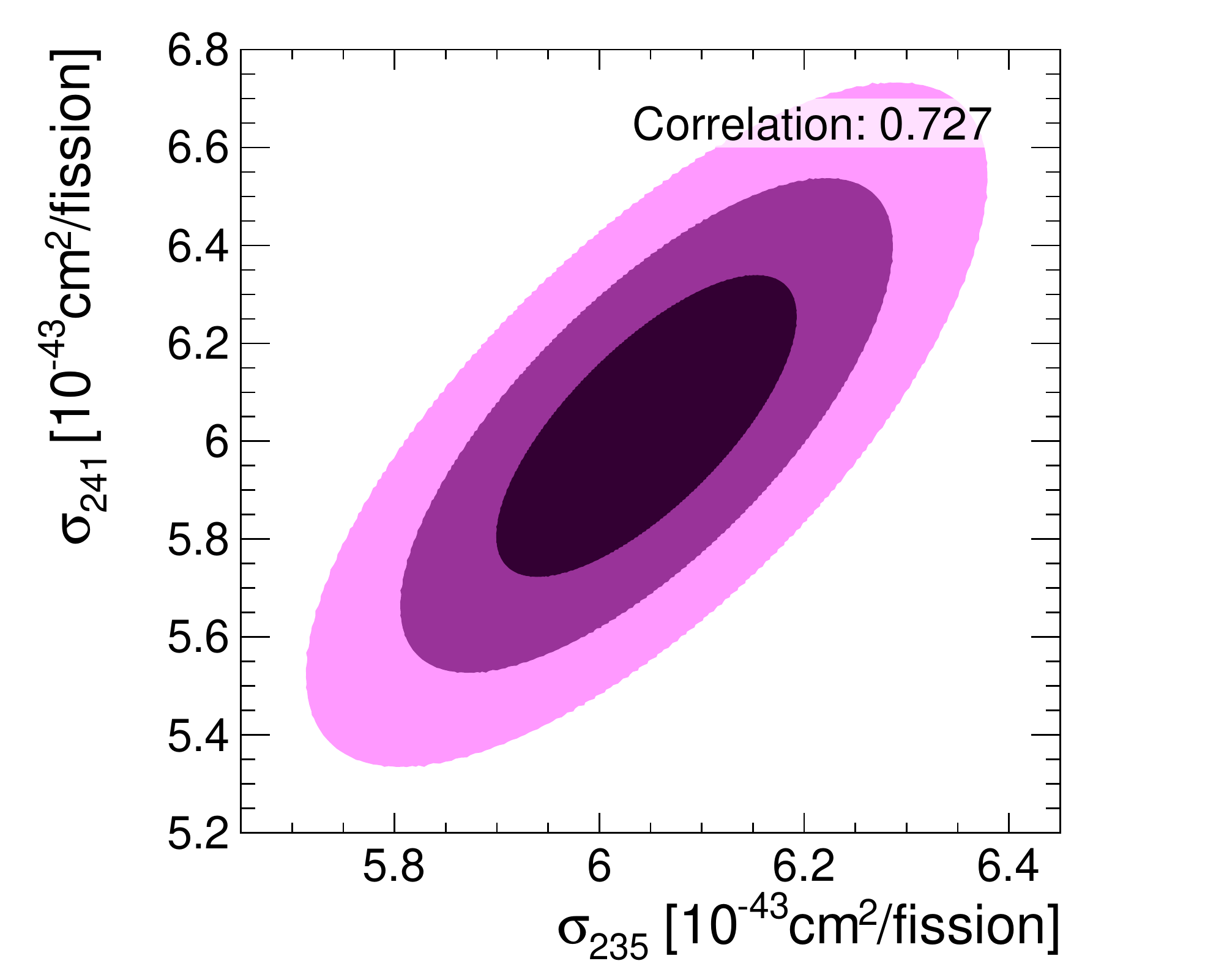}
  \includegraphics[trim={0.6cm 0 0.5cm 0},clip,width=0.33\textwidth]{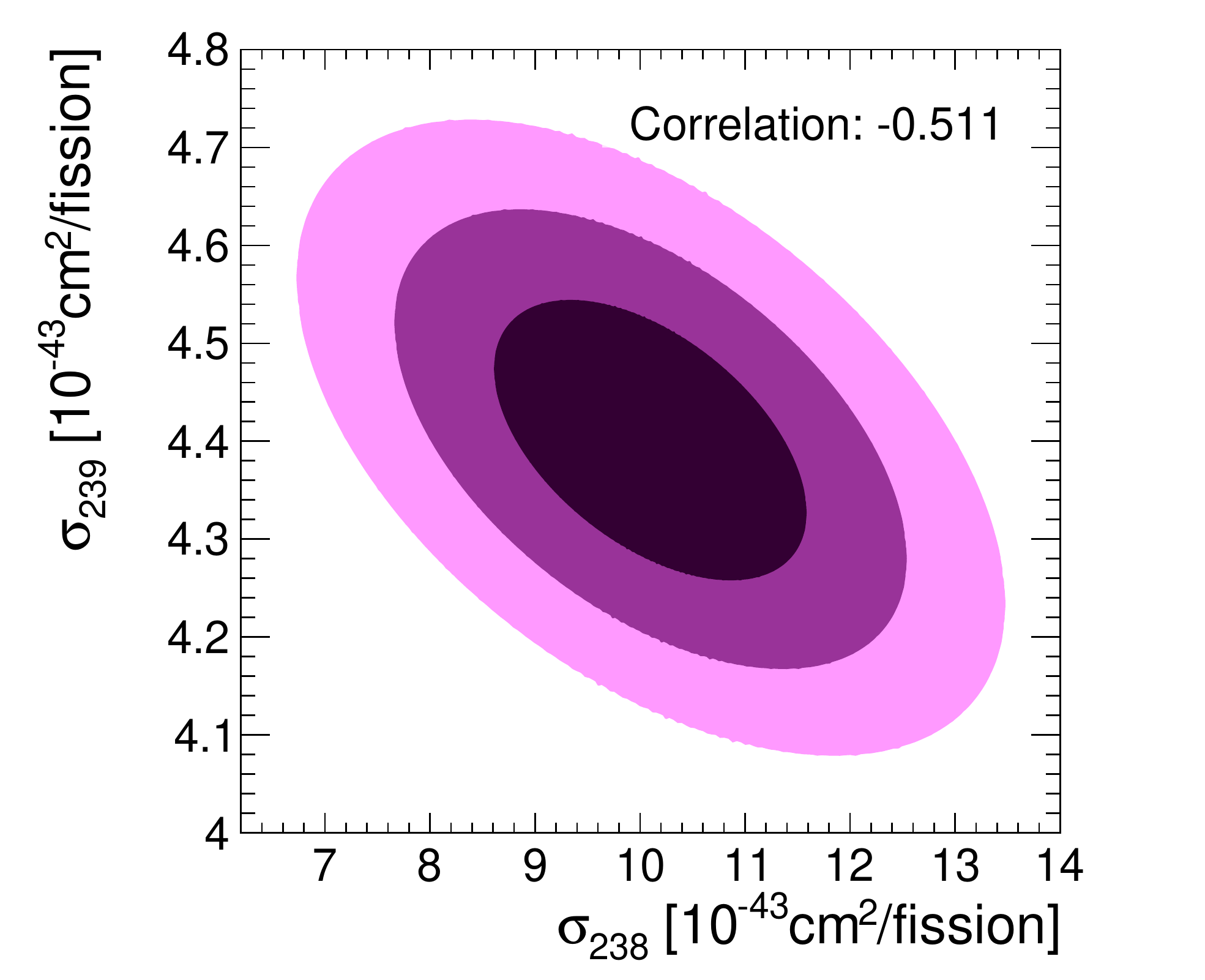}
  \includegraphics[trim={0.6cm 0 0.5cm 0},clip,width=0.33\textwidth]{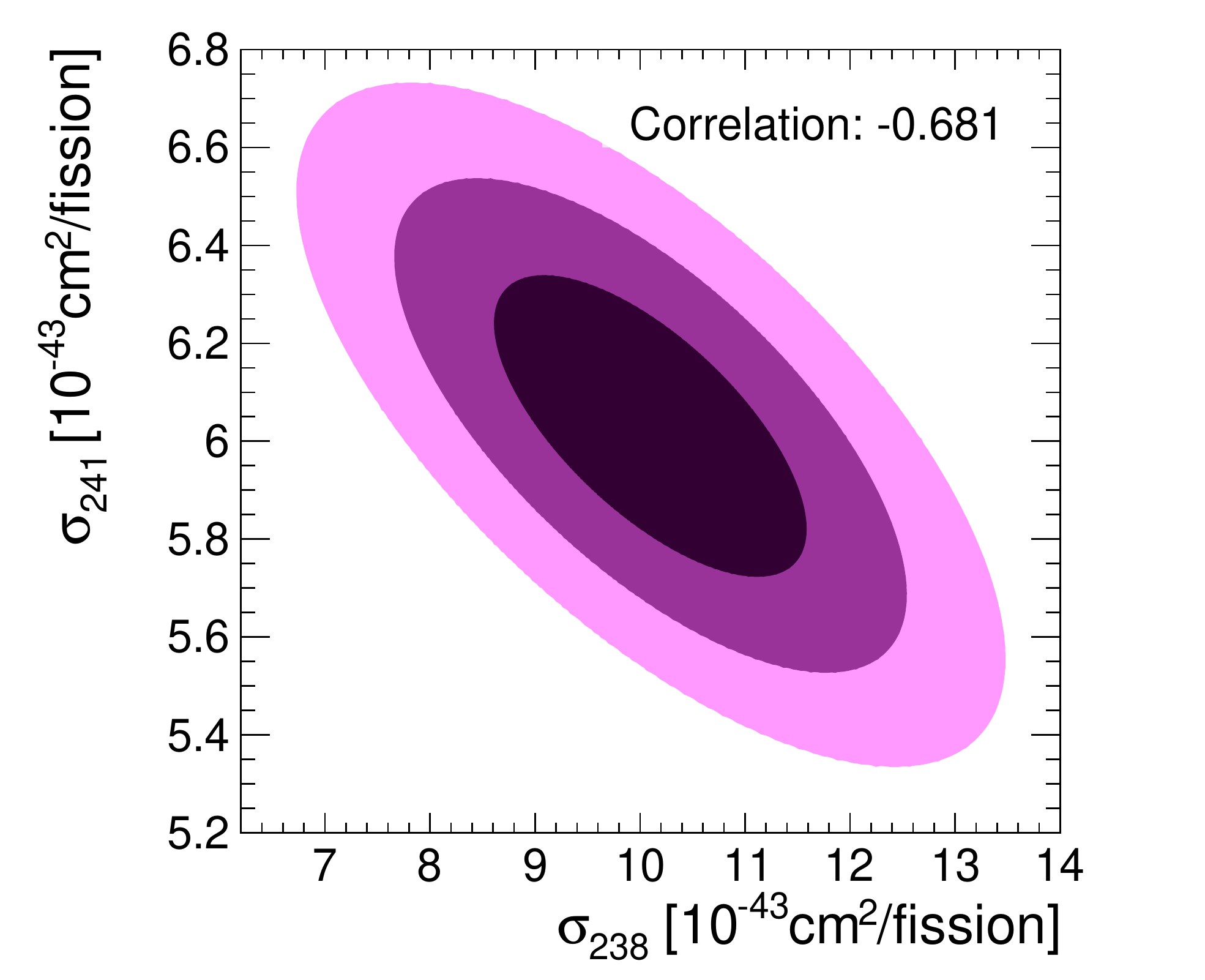}
  \includegraphics[trim={0.6cm 0 0.5cm 0},clip,width=0.33\textwidth]{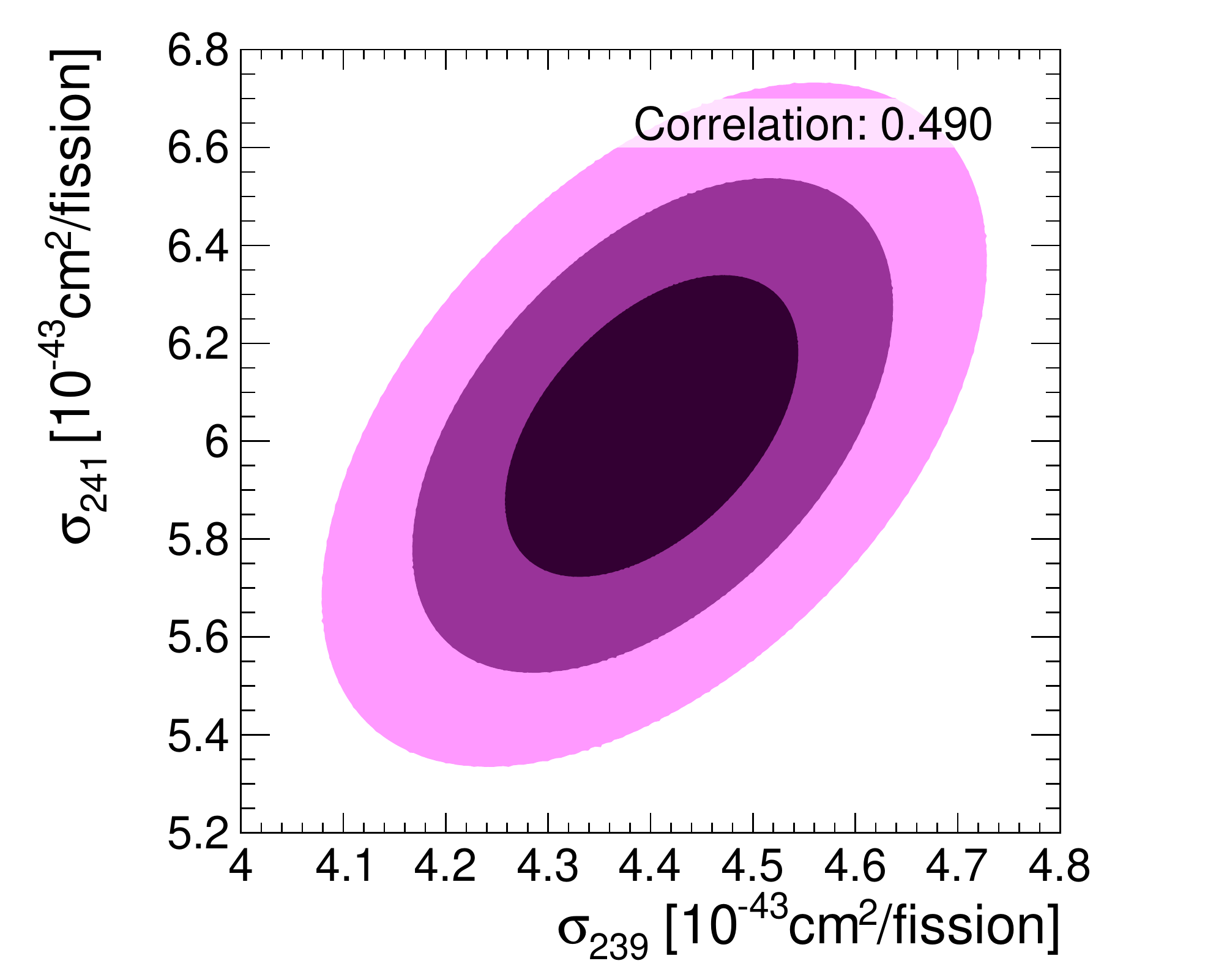}
  \caption{Isotopic IBD yield contours for a combined fit of hypothetical HEU, LEU, and RG-MOX datasets.  In each panel, fits are marginalized over the undepicted isotopes.  Correlation coefficients between each pair of isotopes are provided in the legend.}
  \label{fig:TriangleNew}
\end{figure*}

Reactors burning mixed-oxide (MOX) fuels are another promising venue for performing IBD yield measurements with unique $F_{i}$ combinations.  
In particular, the RG-MOX measurement case may be an imminently realizable one, given the presence and operation of RG-MOX commercial cores in Europe and Japan.  
The 50\% reactor-grade mixed-oxide (RG-MOX) core described in Section~\ref{subsec:future} features $F_{239}$ far higher than an LEU core and broad variations in $F_{241}$ from nearly 15\% at reactor start-up to roughly 25\% after one cycle.  
Ratios $F_{239}$/$F_{241}$ vary much more widely from cycle beginning (27\%) to end (45\%) compared to the LEU reactor case above.  
Amidst these substantial fission fraction variations, \uEight~fractions remain relatively consistent between LEU and RG-MOX cases, offering further opportunity for reduction in degeneracy between \uEight~and the other isotopes.  

Addition of a hypothetical ten-datapoint IBD yield dataset from this RG-MOX reactor core provides substantial enhancements in IBD yield precision when added to those of the short-baseline HEU and LEU datasets, which are also summarized in Table~\ref{tab:Uncertainty_Improvement}.  
Expected precision of yields $\sigma_{239}$~and $\sigma_{241}$~are improved by another factor of $\sim 2$ and $\sim 3$ respectively when the hypothetical RG-MOX is added to the fit alongside the hypothetical HEU and LEU datasets.  
Meanwhile, $\sigma_{238}$ yield precision is also tightened to 9.7\% expected relative uncertainty.  
Correlations between yield fit parameters for this case are also pictured in Figure~\ref{fig:TriangleNew}, and appear further reduced between \pNine~and \pOne~with respect to the hypothetical HEU+LEU case.  
As with the HEU+LEU case, if measurements are performed instead with a MAD-sized 1\,ton detector target, only modest degradation in precision is seen: $\sigma_{235, 238, 239, 241}$ uncertainties shift from (1.6\%, 9.7\%, 2.2\%, 3.4\%) for a 4\,ton target to (1.6\%, 10.3\%, 2.5\%, 3.9\%) for a 1\,ton target.  
uncertainty.
On the other hand, if the correlation between the reactor measurements are removed, or if the chosen optimistic 1\% HEU thermal power uncertainties are increased to the currently-achievable 2\% level, precision in knowledge of the \uEight~and \pOne~yields are reduced---to 14.9\%, 15.4\% and 4.3\%, 5.0\% respectively--- and are moderately worse than the theoretical yields.
Comparing this with the HEU+LEU case where the precision achievable on \uEight ~yield is 11.1\%, the improvement provided by the addition of RG-MOX reactor data doesn't fully compensate for the loss in precision due to the lack of correlation or a reduction in thermal power uncertainty. 

With measurements at three reactor types -- HEU, LEU, and MOX -- with a common detector, direct IBD-based constraints on \nuebar production by the four primary fission isotopes may be expected to rival or exceed the precision of conversion-based predictions.  
Most of these direct isotopic yield uncertainties are also smaller and more well-defined in origin than the $\mathcal{O}$(5\%) uncertainty attributed to summation predictions for these isotopes.  
Thus, with a global HEU+LEU+MOX dataset, one could generate IBD-based reactor \nuebar flux predictions for many existing or future reactor types free from biases known to be present in conversion-predicted models without sacrificing relative model precision.  

Expected isotopic IBD yield measurement precision delivered by instead combining a ten datapoint weapons-grade mixed-oxide (WG-MOX) measurement with the hypothetical HEU and LEU datasets has also been considered. 
IBD yield uncertainties for a HEU+LEU+WG-MOX measurement set are slightly worse than a HEU+LEU+RG-MOX set for $\sigma_{238}$, $\sigma_{239}$, and $\sigma_{241}$ as shown in Table~\ref{tab:Uncertainty_Improvement}.  
Similarity in results between MOX fuel types should not be too surprising, since both WG-MOX and RG-MOX cycles roughly span a $~\sim 16-17\%$ range in $F_{239}$/$F_{241}$ fission fraction ratios.  

It is worth noting that wide variations in $F_{239}$/$F_{241}$ should also expected to be provided by conventional LEU cores burning entirely fresh fuel, such as would occur upon first operation of a new commercial power plant~\footnote{A schedule of forecasted reactor facility start-ups world-wide are given at \href{https://world-nuclear.org/information-library/current-and-future-generation/plans-for-new-reactors-worldwide.aspx}{https://world-nuclear.org/information-library/current-and-future-generation/plans-for-new-reactors-worldwide.aspx}}.  
In this case, $F_{239}$/$F_{241}$ fission fraction ratios should be expected to vary by well over 10\% over course of a fuel cycle~\cite{Anna}.  
Thus, in lieu of MOX-based options, IBD yield measurement regimes including newly started commercial cores likely serve as another promising avenue for producing precise constraints on all main fission isotopes.  

\subsection{Benefits from Fast Reactor Measurements}

Since fast fission cross-sections of many minor actinides -- particularly \pZero~-- are substantially higher than thermal fission cross-sections, fission fractions in the VTR and PFBR fast reactors are substantially different than those of the high-MOX-fraction conventional core configurations described in~\cite{Bernstein:2016ayp}.  
In particular, \pZero~fissions now compose a non-negligible fraction of the total, and, as a result, \pOne~fission fractions are substantially lower.  
The addition of the two fast reactor dataset to the hypothetical HEU and LEU datasets is also summarized in Table~\ref{tab:Uncertainty_Improvement}.  
The most striking product of introducing these datasets to the fit is the potential for setting the first-ever meaningful constraints on \nuebar production by \pZero.  
We find roughly comparable \pZero~yield measurements when either VTR or PFBR are fitted separately with the other datasets.  
Such a measurement could prompt new and deeper study of fission yields and decay data for this minor actinide, which plays a major role in the operation of next-generation fast reactor systems.  
The level of achievable precision in the $\sigma_{240}$ measurement is primarily driven by the precision in understanding the thermal output of these fast reactor cores -- an instrumentation challenge under active investigation in the nuclear engineering community.  

Inclusion of fast reactor datasets generates only minor improvements in the knowledge of $\sigma_i$ for the other primary fission isotopes beyond that achievable with the HEU+LEU measurement scenario. 
While this results primarily from the general lack of knowledge of the value of $\sigma_{240}$, it also highlights the value delivered by multiple highly systematically correlated measurements at differing fuel composition, like that provided by the MOX reactor cases, in contrast to the single measurement provided by the relatively static composition of these fast reactor cores.  
Were $F_{240}$ to evolve in a meaningful way for either core, it is likely that the isotopic IBD yield knowledge delivered by this core would be substantially improved.  

\section{Discussion and Summary}  
\label{sec:summary}

After observing that the current global IBD yield dataset exhibits some capability to constrain antineutrino production by \uFive, \uEight, \pNine, and \pOne, we have investigated how suites of future systematically-correlated measurements at diverse reactor core types can improve knowledge for these and other fission isotopes.  
We have observed that with the simplest combination of correlated HEU and LEU measurements using a PROSPECT-sized or MAD-sized IBD detector, an IBD yield measurement precision of 12\% or better can be achieved for all four fission isotopes.  
With a combination of HEU, LEU, and RG-MOX datasets, all isotopic yields can be directly measured with a precision rivaling or exceeding the precision claimed by conversion-predicted models.  
If measurements of fast reactors are also included in the global dataset, first constraints of order 25\% precision can be placed on antineutrino production by \pZero.  
Beyond future measurements, we also noted other avenues for improving knowledge of isotopic IBD yields with current data: in particular, measurements performed over multiple LEU fuel cycles, such as Daya Bay and DANSS, can benefit from exploiting known variations in \pOne~between cycles. 

\begin{figure*}[tbph!]
  \centering
  \includegraphics[trim={2cm 1cm 0 0},clip, width=0.49\textwidth]{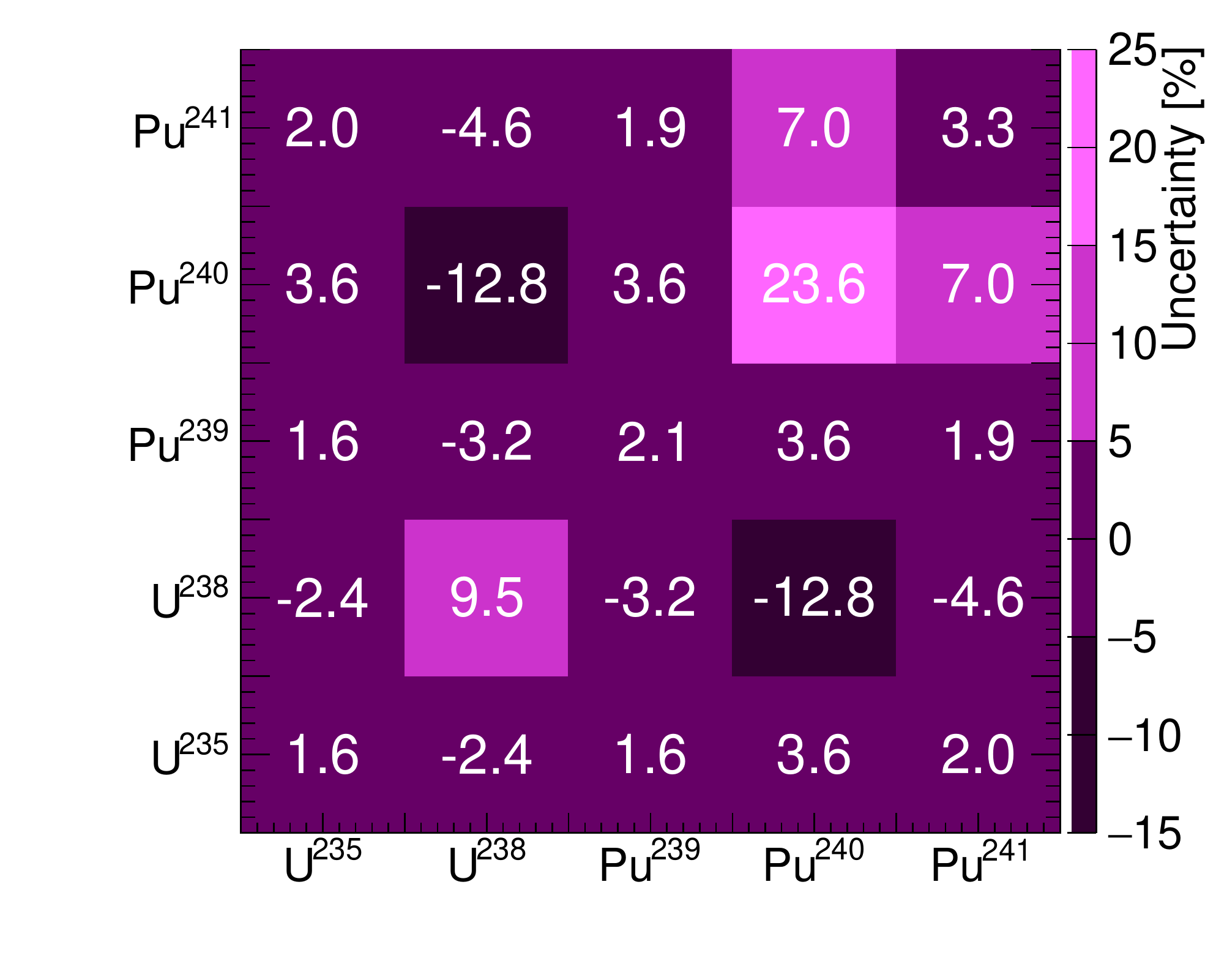}
  \includegraphics[trim={2cm 1cm 0 0},clip,width=0.49\textwidth]{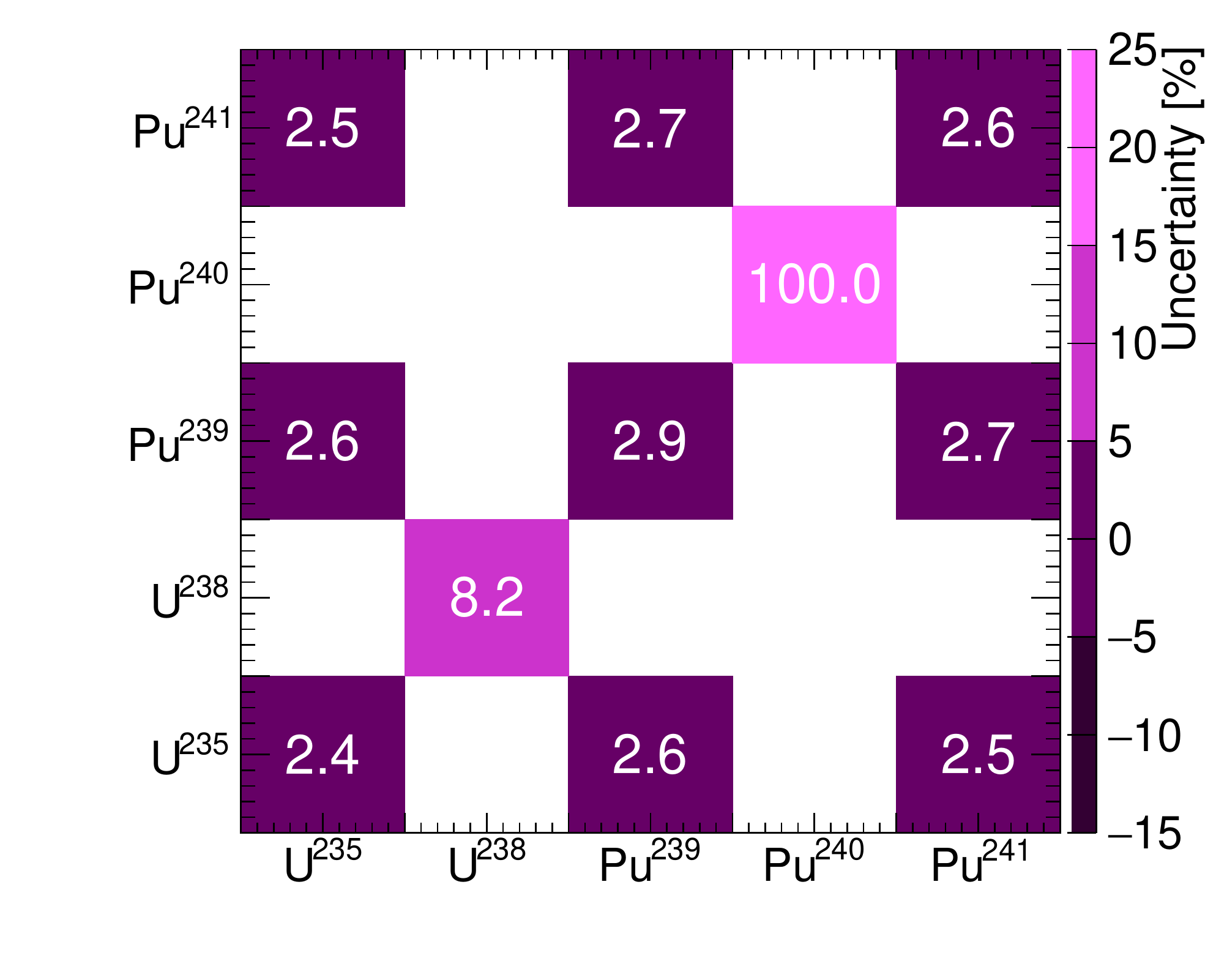}
  \caption{Left: Uncertainties in isotopic IBD yield measurements based on a hypothetical global dataset including HEU, LEU, RG-MOX, and fast reactor IBD yield measurements. Diagonal elements correspond to the uncertainty in isotopic yields given for the ``All" case in Table~\ref{tab:Uncertainty_Improvement}, while off-diagonal elements describe the correlations between them. The values are extracted by taking the square root of the corresponding elements of the correlation matrix and are assigned a negative value where the correlations are negative. Full covariance matrices are provided in the supplementary materials accompanying this paper. Right: Uncertainties in IBD yields predicted by the Huber-Mueller model~\cite{GiuntiGlobal}. Since there are no theoretical models predicting $\sigma_{240}$, we assign 100\% uncertainty on it.}
  \label{fig:UncMatrix}
\end{figure*}

With a combined global dataset in hand from multiple reactor types, one can generate IBD-based reactor \nuebar flux predictions for many existing or future reactor types free from biases known to be present in conversion-predicted models without sacrificing in relative model precision.  
If one considers the full suite of correlated HEU, LEU, RG-MOX and fast reactor measurements (the ``All'' scenario in Table~\ref{tab:Uncertainty_Improvement}), the resultant data-based model would include ($\sigma_{235}, \sigma_{238}, \sigma_{239}, \sigma_{240}, \sigma_{241},$) uncertainties of (1.6, 9.5, 2.1, 23.6, 3.3)\%.  
The correlation between these achievable directly-constrained uncertainties has also been calculated, and can be seen in Figure~\ref{fig:UncMatrix}, alongside those of the Huber-Mueller model~\footnote{Both hypothetical future and Huber-Mueller model IBD yield covariance matrices are included in supplementary materials accompanying this paper.}.  
Besides representing the similar magnitudes in uncertainty, Figure~\ref{fig:UncMatrix} shows direct measurements' reduced correlations between \uFive, \pNine, and \pOne with respect to conversion predictions, which are primarily caused by the common experimental apparatus used at ILL for input fission beta measurements~\cite{bib:ILL_2,bib:ILL_3}.  

This kind of direct and precise understanding of all of the major fission isotopes' contributions to reactor antineutrino emissions would represent movement into an era of `precision flux physics' offering many potential pure and applied physics benefits.  
On the applications side, it would enable unbiased, high-fidelity monitoring, and performing of robust case studies for, a broad array of current and future reactor types.  
Well-measured isotopic antineutrino fluxes could be compared to summation-predicted ones to provide enhanced benchmarking and improvement of nuclear data associated with the main fission isotopes and their daughters, as well as the first meaningful integral datasets for validating the nuclear data of \pZero.  
These models and correlated datasets would allow for precise independent tests of each of the four IBD yield predictions provided by the Huber-Mueller model, enabling thorough investigation of the hypothesis that mis-modelling of one or more isotopes' yields is responsible for the reactor antineutrino anomaly.   
Precise and reliable IBD-based flux constraints would also improve the reach of beyond standard model searches with signal-dominated coherent neutrino-nucleus scattering detectors~\cite{CONNIE:2022hna}.  
Finally, by probing for persistent residual IBD yield deficits common to all isotopes with respect to conversion or summation models, the community can search for enduring hints of sterile neutrino oscillations, even in the presence of other confounding effects, such as neutrino decay or wave packet de-coherence~\cite{Arguelles:2021meu}.  
We encourage the use of the forecasted flux uncertainty matrix provided above and in the supplementary materials as input for future physics sensitivity and use case studies; these exercises would help to directly demonstrate the value of this achievable advance reactor neutrino flux knowledge.  

\section{Acknowledgements}
This work was supported by DOE Office of Science, under award No. DE-SC0008347, as well as by the IIT College of Science.  We thank Anna Erickson, Jon Link, and Patrick Huber for useful comments and discussion, and Nathaniel Bowden and Carlo Giunti for comments on early manuscript drafts.  

\bibliographystyle{apsrev4-1}
\bibliography{main}{}

\end{document}